\documentclass[fleqn,usenatbib]{mnras}
\usepackage{graphicx}	
\usepackage{amsmath}	
\usepackage{amssymb}
\usepackage{subcaption}
\usepackage{multirow}
\usepackage{amsmath}
\usepackage{url}
\usepackage{setspace}
\usepackage{longtable}
\usepackage{pdflscape}
\usepackage{float}
\usepackage{booktabs}
\usepackage{tabularx}
\usepackage{amssymb}
\usepackage{wasysym}
\usepackage{rotating}
\usepackage{adjustbox}
\usepackage{xspace}
\usepackage{xcolor}
\usepackage[T1]{fontenc}
\usepackage{adjustbox}
\usepackage{enumitem,calc}

\newcommand{\kms}{kms$^{-1}\,$}

\newcommand{\hi}{{H\,{\small I}}\xspace }

\newcommand{\perbeam}{beam$^{-1}$}
\newcommand{\simba}{{\sc Simba}}

\newcommand\Tstrut{\rule{0pt}{3ex}}         
\newcommand\Bstrut{\rule[-1.5ex]{0pt}{0pt}}   

\title[The baryonic Tully-Fisher relation over the last billion years]{MIGHTEE-HI: The baryonic Tully-Fisher relation over the last billion years}

\author[A.~A.~Ponomareva et al.]{Anastasia A.~Ponomareva$^1$\thanks{Email: anastasia.ponomareva@physics.ox.ac.uk}, Wanga Mulaudzi$^2$, Natasha Maddox$^3$, Bradley S.~Frank$^{4,5,2}$,
\newauthor Matt J.~Jarvis$^{1,6}$, Enrico M. Di Teodoro$^{7,8}$, Marcin Glowacki$^{9,6}$, Ren\'{e}e C. Kraan-Korteweg$^{2}$,
\newauthor Tom A. Oosterloo$^{10,11}$, Elizabeth A. K. Adams$^{10,11}$, Hengxing Pan$^{6,1}$, Isabella Prandoni$^{12}$, 
\newauthor Sambatriniaina H. A. Rajohnson$^2$, Francesco Sinigaglia$^{13,14}$, Nathan J.~Adams$^1$, Ian Heywood$^{1,15,4}$,
\newauthor Rebecca A. A. Bowler$^1$, Peter W. Hatfield$^1$, Jordan D. Collier$^{5,16,17}$ and Srikrishna Sekhar$^{5,18,6}$\\
$^1$Oxford Astrophysics, Denys Wilkinson Building, University of Oxford, Keble Rd, Oxford, OX1 3RH, UK\\
$^2$Department of Astronomy, University of Cape Town, Private Bag X3, Rondebosch 7701, South Africa \\
$^3$Faculty of Physics, Ludwig-Maximilians-Universit{\"a}t, Scheinerstr. 1, 81679 Munich, Germany\\
$^4$South African Radio Astronomy Observatory, 2 Fir Street, Observatory, 7925, South Africa\\
$^5$The Inter-University Institute for Data Intensive Astronomy (IDIA), and University of Cape Town, \\
Private Bag X3, Rondebosch, 7701, South Africa\\
$^6$Department of Physics and Astronomy, University of the Western Cape, Robert Sobukwe Road, Bellville 7535, South Africa\\
$^7$Department of Physics \& Astronomy, Johns Hopkins University, Baltimore, MD 21218, USA\\
$^8$Space Telescope Science Institute, 3700 San Martin Drive, Baltimore, MD 21218, USA\\
$^9$Inter-University Institute for Data Intensive Astronomy, Bellville 7535, South Africa\\ 
$^{10}$ASTRON, the Netherlands Institute for Radio Astronomy, Oude Hoogeveesedijk 4,7991 PD Dwingeloo, The Netherlands\\
$^{11}$Kapteyn Astronomical Institute, PO Box 800, 9700 AV Groningen, The Netherlands\\
$^{12}$INAF-IRA, Via P. Gobetti 101, 40129, Italy\\
$^{13}$Department of Physics and Astronomy, Universit\'{a} degli Studi di Padova, Vicolo dell'Osservatorio 3, I-35122, Padova, Italy\\
$^{14}$INAF - Osservatorio Astronomico di Padova, Vicolo dell'Osservatorio 5, I-35122, Padova, Italy \\
$^{15}$Department of Physics and Electronics, Rhodes University, PO Box 94, Makhanda, 6140, South Africa\\
$^{16}$School of Science, Western Sydney University, Locked Bag 1797, Penrith, NSW 2751, Australia\\
$^{17}$CSIRO Astronomy and Space Science, PO Box 1130, Bentley, WA, 6102, Australia\\
$^{18}$National Radio Astronomy Observatory, 1003 Lopezville Road, Socorro, NM 87801, USA\\}

\date{Accepted 2021 September 10. Received 2021 August 13; in original form 2021 June 25}

\pubyear{2021}

\begin{document}
\label{firstpage}
\pagerange{\pageref{firstpage}--\pageref{lastpage}}
\maketitle

\begin{abstract}
Using a sample of 67 galaxies from the MIGHTEE Survey Early Science data we study the \hi-based baryonic Tully-Fisher relation (bTFr), covering a period of $\sim$one billion years ($0 \leq z \leq 0.081 $). We consider the bTFr based on two different rotational velocity measures: the width of the global \hi profile and $\rm V_{out}$, measured as the outermost rotational velocity from the resolved \hi rotation curves. 
Both relations exhibit very low intrinsic scatter orthogonal to the best-fit relation ($\sigma_{\perp}=0.07\pm0.01$), comparable to the {\it SPARC} sample at $z \simeq 0$. The slopes of the relations are similar and consistent with the $ z \simeq 0$ studies ($3.66^{+0.35}_{-0.29}$ for $\rm W_{50}$ and $3.47^{+0.37}_{-0.30}$ for $\rm V_{out}$). We find no evidence that the bTFr has evolved over the last billion years, and all galaxies in our sample are consistent with the same relation independent of redshift and the rotational velocity measure. Our results set up a reference for all future studies of the \hi-based bTFr as a function of redshift that will be conducted with the ongoing deep SKA pathfinders surveys. 

\end{abstract}


\section{Introduction}
\label{sec:intro}
The Tully-Fisher relation (TFr, \citealt{tf77}) is among the most fundamental dynamical scaling relations for spiral galaxies. It links the luminosity of a spiral galaxy to its rotational velocity through a very tight correlation. It was first used as a redshift independent tool to measure distances to galaxies, with the aim to measure the peculiar velocities and reconstruct local galaxy flows (e.g. \citealt{courtois2012, tully2013, tully2014,cosmic3, cosmic4}). Since then, it has been extensively studied in the field of galaxy formation and evolution. 
It has been shown to hold for rotating galaxies of all morphological types \citep{Chung2002, courteau03, Heijer2015, Karachentsev2016}, in different environments \citep{willick99, Abril-Melgarejo2021}, and over a large wavelength range from FUV to NIR \citep{mverheijen2001,ponomareva2017}.
Consequently, it has become a major tool with which to test galaxy formation and evolution models by examining their ability to reproduce the statistical properties of the TFr (slope, scatter and zero point). 

The nature of the TFr is considered to be understood as the relation between two fundamental properties of spiral galaxies: their baryonic content, characterised by the luminosity, and the total dynamical mass, characterised by the rotational velocity. The luminosity traces the stellar mass of a galaxy, which is the reason why the infrared TFr has been the preferred choice for distance measurement, since the infrared is emitted predominantly from the old stellar population responsible for the bulk of the stellar mass \citep{sorce13}. The stellar mass, in turn, is a good proxy for the total baryonic mass. However, this is a reasonable assumption only for galaxies with rotational velocities $\gtrsim 100$ \kms. The lower velocity regime is populated by the gas-rich dwarf galaxies whose mass is dominated by cold gas (\hi) rather than stars \citep{mcgaugh00}. Thus, inclusion of the cold gas mass resulted in the most fundamental form of the TFr: the baryonic Tully-Fisher relation (bTFr), the tight linear relation that spans $\sim 5$ dex in baryonic mass \citep{mcgaugh12, lelli2016, lelli2019}.

At $z \simeq 0$ the bTFR has been extensively studied for various galaxy samples, and has been the focus for testing different methods to evaluate the stellar mass and the rotational velocity of galaxies. The largest study to date is based on the {\it SPARC} (\emph{Spitzer} Photometry \& Accurate Rotation Curves) database, which consists of 175 spiral galaxies with high-quality \hi rotation curves \citep{lelli2016b}. 
The importance of the resolved \hi rotation curves for the bTFr studies have been extensively discussed in the literature \citep{mverheijen2001, noordermeer2007, lelli2016, ponomareva2017, ponomareva2018}. \citet{lelli2019} analysed the {\it SPARC} sample based on different velocity definitions, such as
$\rm W_{50}$ - the most commonly used rotational velocity measure derived from the integrated \hi line profile,
$\rm V_{flat}$ - velocity measured at the flat part of the extended \hi rotation curve, and
$\rm V_{max}$ - the maximum rotational velocity measured from the \hi rotation curve. They found that the use of $\rm V_{flat}$ yields the tightest bTFr with the steepest slope. These results agree with previous studies, which were based on much smaller samples \citep{mverheijen2001,ponomareva2018}, as well as with the recent studies based on simulations \citep{Glowacki2020a}. The results of the {\it SPARC} study provide important constraints on theories of galaxy formation and evolution: (1) the intrinsic scatter of the bTFr at $z \simeq 0$ is below the lowest value expected in $\Lambda$CDM cosmology \citep{dutton2012}; (2) the bTFr slope, when based on $\rm V_{flat}$, must be in the range of 3.5 to 4, which is higher than the slope predicted by the basic $\Lambda$CDM models \citep{mcgaugh12}; (3) the bTFr residuals do not correlate with galaxy size or surface brightness, contrary to the expectations from galaxy formation and evolution models \citep{desmond2015}; (4) \citet{iorio2017} showed that there is no evidence for curvature at the low-mass end of the bTFr, despite the predictions by some semi-analytical galaxy formation models \citep{TG11, desmond12}. Moreover, \citet{pavel2019,pavel2020} have recently discovered that ultra-diffuse galaxies seem to be the only known population of galaxies which lies off the bTFr. This result challenges our current understanding of the feedback processes in dwarf galaxies.


To date, \hi remains difficult to detect in emission beyond $z \simeq 0$, particularly for radio interferometers which are needed to provide spatially resolved \hi kinematics. Therefore, other kinematic tracers of galaxies have been used to study the potential evolution of the bTFr with redshift. \citet{topal2018} used carbon monoxide (CO) and found no bTFr evolution over the redshift range $\rm 0.05 \leq z \leq 0.3$. \citet{ubler2017} and \citet{tiley2016,tiley19} used optical kinematic tracers such as H$\alpha$ emission line and found conflicting results at $z \approx 1$. The use of these kinematic tracers, however, has various drawbacks. First, both CO and optical emission lines have compact distributions and may not fully probe the Dark Matter (DM) halo potential, unlike \hi, which extends far beyond the optical radius \citep{frank2016}. Second, it is extremely challenging to compare the results at higher redshift to those at $z = 0$. For example, \citet{tiley19} showed that observational data quality can strongly bias the statistical properties of the measured TFr. This is especially important when comparing data that are not produced and analysed homogeneously. Ideally one would need to compare the statistical properties of the bTFr at different redshifts using a blind, volume-limited survey of galaxies, selected in the same way. 

The predictions from cosmological simulations and semi-analytical models of galaxy formation regarding the evolution of the bTFr are also limited. 
A recent study by \citet{Glowacki2020b} aimed to provide predictions for future \hi surveys using the state-of-the-art cosmological simulation \simba\ \citep{dave2019} and found a clear evolution of the best-fit linear parameters of the bTFR over the redshift range $z = 0\to1$, which can be mostly explained by the differences in the merger histories of the DM haloes.

Fortunately, the forthcoming \hi surveys with the SKA pathfinder telescopes, such as LADUMA (Looking At the Distant Universe with the MeerKAT Array, \citealt{blyth2016}) and DINGO (the Deep Investigation of Neutral Gas Origins, \citealt{meyer2009}), together with existing surveys such as CHILES (COSMOS \hi Large Extragalactic Survey, \citealt{chiles2019}) and BUDHIES (Blind Ultra-Deep \hi Environmental Survey, \citealt{budhies2020}), have the potential to systematically study \hi in galaxies over a large range of redshifts. Another such survey is MIGHTEE (the MeerKAT International GigaHertz Tiered Extragalactic Exploration), one of the first deep, blind, medium-wide interferometric surveys for \hi ever undertaken \citep{jarvis2016}. It will detect more than 1000 galaxies in \hi up to $z = 0.6$, thus allowing the systematic study of the evolution of the neutral gas content of galaxies over the past 5 billion years in different environments \citep{ranchod2021} using direct detections and statistical stacking methods (\citealt{maddox2021, Pan2020, Pan2021}). 

In this work we use the MIGHTEE Early Science data to perform, for the first time, a homogeneous study of the \hi-based bTFr over the last billion years ($\rm 0 \leq z \leq 0.081$). Furthermore, we consider the bTFr based on two velocity measures: $\rm W_{50}$ from the corrected width of the global \hi profile, and $\rm V_{out}$, the rotational velocity measured at the outermost point of the resolved \hi rotation curves. This allows us to study how the statistical properties of the bTFr change with redshift and with different definitions of the rotational velocity. Moreover, this is the first study which tests a completely automated version of $\rm^{3D}$Barolo \citep{3dbarolo} at higher redshift, software that was developed to derive \hi rotation curves for the marginally-resolved galaxies, in preparation for the new generation of large \hi surveys.

This paper is organised as follows. Section \ref{sec:survey} describes the MIGHTEE Survey and the Early Science data. Section \ref{sec:mass} describes the baryonic mass measurements. Section \ref{sec:velocity} describes different velocity measurements. Section \ref{sec:results} discusses the results. Summary and conclusions are presented in Section \ref{sec:summary}. 

\section{MIGHTEE survey}
\label{sec:survey}
The MIGHTEE is a survey of four well-known deep, extragalactic fields currently being observed by MeerKAT, the SKA precursor radio interferometer located in South Africa \citep{Jonas_2009}. MeerKAT consists of 64 offset Gregorian dishes (13.5 m diameter main reflector and 3.8 m sub-reflector), and equipped with three receivers: UHF--band ($580 < \nu < 1015$ MHz), L--band ($900 < \nu < 1670$ MHz) and S--band ($1750 < \nu < 3500$ MHz). The MeerKAT data are collected in spectral mode, which makes MIGHTEE a spectral line, continuum and polarisation survey. The \hi emission project within the MIGHTEE survey (MIGHTEE--\hi) is described in detail in \citet{maddox2021}. 

The Early Science MIGHTEE--\hi observations were conducted between mid-2018 and mid-2019. These observations were performed with the full array (64 dishes) in L--band, but with a limited spectral resolution (208 kHz, 44 \kms at $z=0$). The MIGHTEE--HI Early Science visibilities were processed with the {\sc ProcessMeerKAT} calibration pipeline (Frank et al. in prep). The pipeline is {\sc Casa}\footnote{\href{http://casa.nrao.edu}{http://casa.nrao.edu}}-based \citep{McMullin2007} and performs standard data reduction and calibration tasks for spectral line data such as flagging, bandpass and complex gain calibration. Spectral line imaging was performed using {\sc Casa}'s task \texttt{TCLEAN}  ({\sc robust=0.5}). The continuum subtraction was done in both the visibilities and imaging domain using standard {\sc Casa} routines  \texttt{UVSUB} and \texttt{UVCONTSUB}. Per-pixel median filtering was applied to the resulting data cubes to reduce the impact of the direction-dependent artefacts. A full description of the data reduction strategy and data quality assessment will be presented in Frank et al. (in prep). The Early Science data used in this paper is summarized in Table \ref{tbl_data}. 

The source finding was performed visually, using {\sc CARTA} (The Cube Analysis and Rendering Tool for Astronomy, \citealt{carta}), and unguided by the deep optical information available for these well-studied fields, by the MIGHTEE-\hi group. The total Early Science sample consists of 276 objects each with an identified optical counterpart \citep{maddox2021}. 

\begin{table}
\centering
\begin{tabular}{ll}
\hline
\hline
Area covered & 1 deg$^2$ COSMOS field\Tstrut\\
& 3 deg$^2$ XMMLSS field \\
Obs. time & 16h COSMOS field\\
& 14h each 1 deg$^2$ of XMMLSS field  \\ 
Frequency range& $1310-1420$ MHz\\
Channel width                            & 208 kHz               \\
Pixel size                       & 2"                                      \\
Median \hi channel rms noise & 85 $\mu$Jy \perbeam               \\
Synthesised beam   & 14.5" $\times$ 11"    COSMOS field                              \\
           & 12" $\times$ 10"    XMMLSS field                                              \\
$N_\mathrm{HI}$ sensitivity ($3\sigma$)  & 1.6 $\times \, 10^{20}\, \mathrm{cm}^{-2}$ (per channel)                \Bstrut\\ 
\hline           
\hline
\end{tabular}
\caption{Brief description of the MIGHTEE-\hi Early Science data used in this paper.}
\label{tbl_data}
\end{table}

\section{Baryonic mass}
\label{sec:mass}
Throughout this paper we refer to the baryonic mass (M$_{\rm bar}$) as a sum of the stellar mass component and the total neutral atomic gas mass: 
$\rm M_{gas} = M_{\hi} \times 1.4$, where $\rm M_{\hi}$ is the \hi gas mass, and the factor $1.4$ accounts for the primordial abundance of helium and metals \citep{arnett1999}. 
We do not take the molecular gas into account, since it has been found to have a negligible effect on the statistical properties of the bTFr \citep{ponomareva2018}. 

\subsection{Stellar mass}
\label{stellar_mass}
All MIGHTEE targeted fields have a comprehensive range (from X-ray to far-infrared) of ancillary data from various multi-wavelength photometric and spectroscopic surveys.
The main photometric surveys, relevant for the stellar mass measurements, include the optical photometry from Canada-France-Hawaii Telescope Legacy Survey (CFHTLS, \citealt{Cuillandre2012}), HyperSuprimeCam (HSC, \citealt{Aihara2018, Aihara2019}), near-infrared photometry
from the VISTA Deep Extragalactic Observations (VIDEO, \citealt{jarvis2013}) and UltraVISTA \citep{McCracken2012}. All ancillary data are described in detail in \citet{maddox2021}. 

The magnitudes of the sample galaxies were measured by extracting the flux within an elliptical aperture, which was defined in the $g$--band and applied to the 
$urizYJHK_{s}$--bands. The Spectral Energy Distribution (SED) fitting code LePhare \citep{Arnouts1999, Ilbert2006} was then used to derive the stellar properties of the galaxies, such as stellar mass, stellar age and star-formation rate. 
We adopt a conservative uncertainty of the stellar mass for each galaxy of $\sim 0.1$ dex \citep{adams2021}.

\subsection{\hi Mass}
The total \hi mass of each galaxy was calculated using:
\begin{equation}
\left(\frac{M_{\mathrm{HI}}}{M_{\odot}}\right) = \frac{2.356 \times 10^{5}}{1+ {z}}\left(\frac{D_{L}}{\mathrm{Mpc}}\right)^{2}\left(\frac{S}{\mathrm{Jy}\,\mathrm{km\,s}^{-1}}\right),
\end{equation}
where $D_L$ is the cosmological luminosity distance to the source, $z$ is redshift and $S$ is the integrated \hi flux density, calculated from the moment-0 maps, as described in \citet{meyer2017}. Moment-0 maps were constructed for each galaxy individually as follows: first, the cubelets (cutouts from the original data cube centred on each detection) were smoothed to a circular beam of $20" \times 20"$ and clipped at 3$\sigma$ threshold (where $\sigma$ was obtained by measuring the noise over an emission-free region of the cubelet and calculating the standard deviation). The resulting mask was applied to the original resolution cubelet, thus allowing to take into account low column density diffuse \hi emission. The moment-0 maps were constructed using the resulting masked cubelet \footnote{Throughout the paper we use the \texttt{spectral-cube} package \citep{speccube} for our \hi data analyses, if not stated otherwise.}. Then, every moment-0 map was examined by eye, and emission from the galaxy was isolated by masking out the noise peaks and negative flux values if any were present. 
To calculate the error on the integrated flux $S$, we projected the source mask, used to construct the moment-0 map, to four emission free regions around the detection.
We then measured the signal in each of the four regions and defined
the uncertainty in the integrated flux of a galaxy as the mean rms scatter of the four
flux measurements in the projected masks \citep{mpati2016}.
As a result, the typical uncertainty on the \hi mass varies from $\sim 5\%$ for the high-mass galaxies to $\sim 20\%$ for the lowest mass objects ($\rm M_{\hi} \leq 10^{8}M_{\odot}$). 

\section{Rotational velocities}
\label{sec:velocity}
\subsection{Inclinations}
\label{sec:incl}
The observed rotational velocity of a galaxy can be converted to an intrinsic velocity by taking into proper account the geometry of the source. Thus, any rotational velocity measure should be corrected for the inclination effect. For face-on discs inclination corrections become very large due to the $\sin(i)$ dependence. Even though there are proposed methods to estimate the statistical properties of the bTFr with no prior knowledge of the inclination \citep{obreschkow2013}, they assume that galaxies follow the known functional form of the relation and a galaxy sample obeys a TFr with normal scatter, which might not be the case for a $z > 0$ study.

Usually inclinations are measured using infrared photometry due to low extinction in the infrared bands. However, infrared bands trace older stellar populations which do not reside in the thin disk. A parameter that accounts for the thickness of a galactic disk ($q0$) is then used in addition to the axis ratio to measure the inclination \citep{mihalas1981}. To date, there is no agreement on the best value for $q0$. Some studies argue that it depends on the morphology of a galaxy and should vary \citep{giovanelli1997}, while \citet{tully2009} suggest that $q0$ should be fixed for all galaxies to avoid systematic uncertainties. 

Conveniently, in addition to the stellar disk, also the \hi disks can be used to measure the inclination angles. In general, the \hi disk is much thinner than the stellar disk and its intrinsic thickness can be neglected \citep{Verheijen2001}. However, due to disk flaring the \hi disk can become significantly thicker in the outer parts \citep{Bacchini2019}. We tested a thicker disk by setting $q0=0.2$, but found consistent results. Therefore, in what follows we assume an infinitely thin \hi disk for simplicity and for fair comparison with the measurements from $\rm^{3D}$Barolo (see Section \ref{subsec:incl}).

We measure inclination angles ($i_{\hi}$) of our sample galaxies using \hi moment-0 maps as:
\begin{equation}
\cos^2(i_{\hi})=\frac{b^2-\theta_{b}^2}{a^2-\theta_{a}^2},
\end{equation}
where $b$ and $a$ are the minor and major axis of the \hi moment-0 map, measured by fitting an ellipse to the outermost reliable contour equal to $\rm 1 \, M_{\odot}/pc^{2}$, $\theta_{b}$ and $\theta_{a}$ are the sizes of the synthesised beam, used to correct for the beam smearing effect, which can make galaxies look rounder if they are not well resolved \citep{Verheijen2001}. We assign a conservative error on the disk ellipticity of $\sim 10 \%$ to account for the resolution and disk flaring effects, which results in the mean uncertainty of the $i_{\hi}$ of $\sim 5^{\circ}$.

\begin{figure} 
\centering
\includegraphics[scale=0.6]{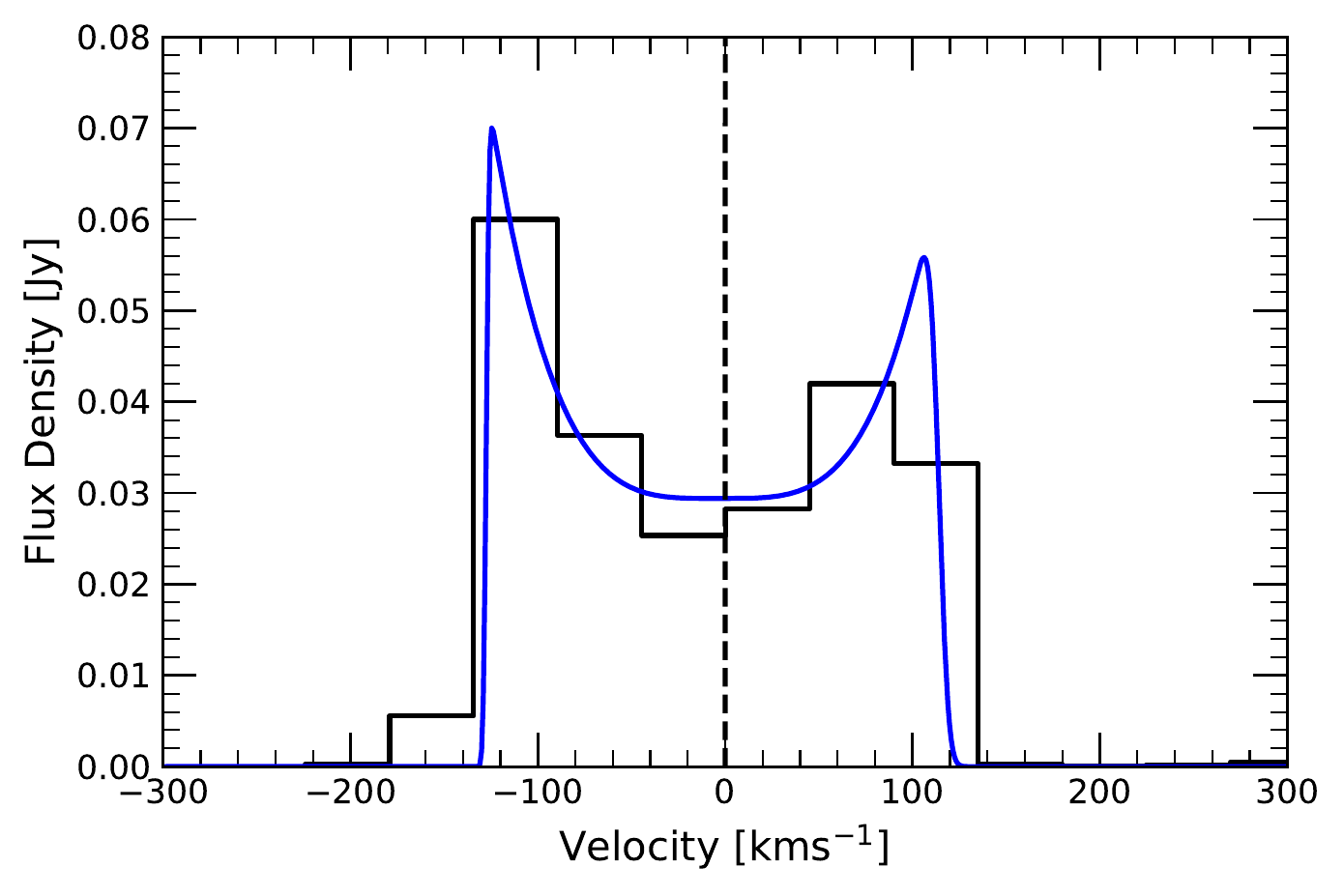}
\caption{The measured global \hi line profile of a galaxy from our sample is shown with the black solid line. The resulting BF \textsc{PyMultiNest} model is shown with the blue solid line.}
\label{fig_W50}
\end{figure}
\begin{figure*} 
\centering
   \includegraphics[scale=0.6]{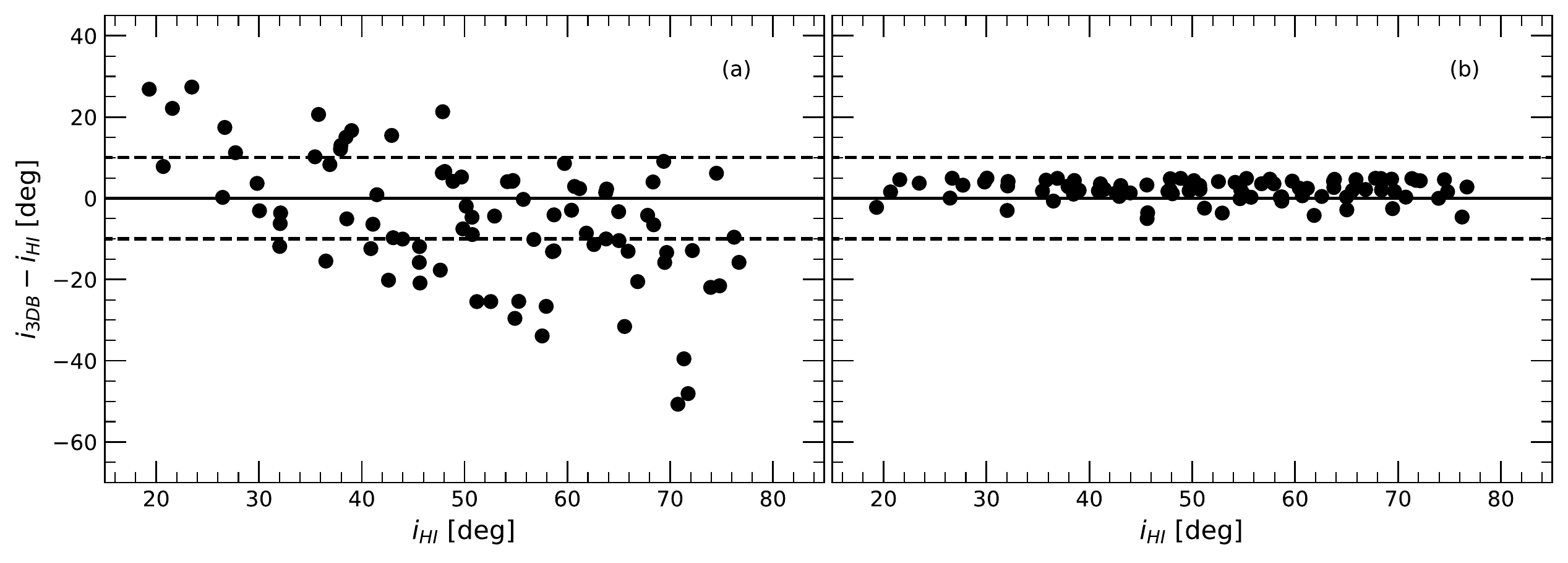}
   \caption{Difference between the inclinations obtained with $\rm^{3D}$Barolo and the \hi inclinations: panel (a) shows the results of a blind, completely automated run; panel (b) shows the results obtained when \hi inclinations were used as initial estimate and kept free.}
\label{fig_incl3db}
\end{figure*}

\begin{figure} 
\centering
   \includegraphics[scale=0.6]{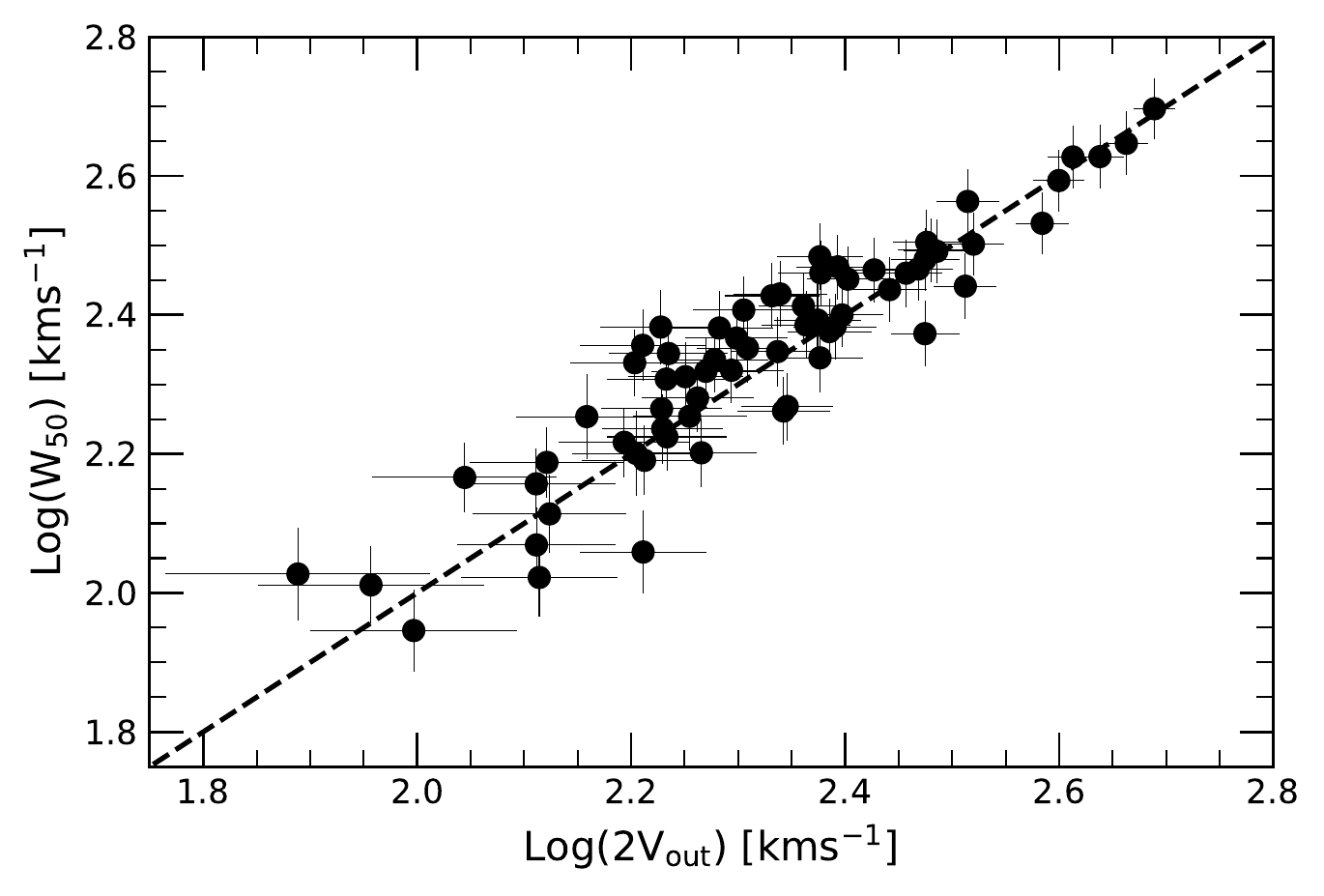}
   \caption{Comparison between corrected $\rm W_{50}$s and rotational velocities measured from the rotation curves. The one-to-one relation is shown by the dashed line.}
\label{fig_W50-Vout}
\end{figure}

\subsection{Line width measurement}
Global \hi line profiles not only hold information about the amount of \hi gas in galaxies, but also about its kinematics.
Usually the line width for the bTFr studies is measured at 50\% of the peak flux density of the global \hi line profile ($W_{50}$) and, if corrected for instrumental broadening and random motions, gives a good representation of the maximum rotational velocity measured from a spatially resolved rotation curve: ${\rm 2V_{max} = W_{50}}/\sin(i)$ (see Figure 6 in \citealt{ponomareva2016}). 

For this study, we measure $\rm W_{50}$ for each galaxy by fitting the Busy Function (BF, \citealt{Westmeier2013}), using multinest to explore the posterior distribution \citep{Feroz2008, Feroz2009}. 

The Busy Function is defined as:
\begin{multline}
B(x) = \frac{a}{4}[\text{erf}(b_1\{{\rm W}+x-x_e\})+1]\times\\ 
[\text{erf}(b_2\{{\rm W}-x+x_e\})+1]\times [c|x-x_p|^n+1], 
\end{multline}
where $\text{erf}(x) = \frac{2}{\sqrt{\pi}}\int^x_0\text{exp}(-t^2)\text{d}t$ is the Gaussian error function, $a$ is the total amplitude scaling factor, $b_1$ and $b_2$ describe the steepness of the line flanks, $\rm W$ is the half-width of the \hi global profile, $x_e$ and $x_p$ are the offsets for the error functions and the polynomial, $c$ is the scaling factor of polynomial trough, and $n$ is the order of the polynomial. 

The fit was performed with \textsc{PyMultiNest} and uses the default initial parameters, such as tolerance=0.5 and live points=1000 \citep{Buchner2014}. The prior distributions of each parameter of the BF were set in the following ranges: $a$: U$\in[0,1]$, $b_1$: U$\in [0,1]$, $b_2$: U$\in [0,1]$, ${\rm W}$: log$\in [1,10^{3}]$, $x_e$: U$\in [-300,300]$, $x_p$: U$\in [-300,300]$, $c$: log$\in [10^{-9},10^{-7}]$, $n$: U$\in [2,8]$, where U stands for the uniform range and log for logarithmic.
Full details of the line-profile fitting will be presented in Mulaudzi et al. (in prep). Figure~\ref{fig_W50} illustrates an example of the fit for one of the sample galaxies. 

The resulting $\rm W_{50}$ values were corrected for instrumental broadening and turbulent motions following the standard prescriptions \citep{Verheijen2001}. Specifically, the instrumental broadening correction for our data is $\sim 26$ \kms, while $5$ \kms was adopted as a standard correction for the turbulent motions when $\rm W_{50}$ is matched to $\rm V_{max}$ \citep{ponomareva2016}. 

The posterior distributions sampled by \textsc{PyMultiNest} provide the typical error associated with $\rm W_{50}$ to be $\sim 10$ \kms. We combine this error in quadrature with a systematic uncertainty of $10\%$ of the line width to account for the masking applied prior to the fit, which is not fully accounted for in the fitting process.

\subsection{3D kinematic modelling}
Though $\rm W_{50}$ can provide a reliable estimate of the rotational velocity of a galaxy, its measurement is dominated by the high-density gas which tends to reside in the central regions of a galaxy. The kinematics of this gas may not be representative of the full potential of the system and $\rm W_{50}$ may differ from the velocities measured in the outer parts of spatially resolved rotation curves \citep{mverheijen2001, ponomareva2016, lelli2019}. 
Therefore, to be able to study the bTFr in detail and compare to the existing studies at $z=0$  accurate 3D kinematic modelling is required.

Kinematic modelling has always been a complex procedure. The so-called tilted-ring modelling technique \citep{rogstad1974} originally developed to fit the 2D velocity fields of spiral galaxies, required \hi observations to have very high spatial and velocity resolution, coupled with high signal-to-noise ratio (SNR) to derive a good-quality rotation curve. However, with new observational facilities and in preparation for the modern \hi surveys, new techniques and software have been developed that allow high quality kinematic modelling, even for marginally 
resolved galaxies, exploiting the full 3D parameter space of the data cubes (e.g. $\rm^{3D}$Barolo, \citealt{3dbarolo}, TiRiFiC, \citealt {tirific}).

Nowadays kinematic modelling software is able to constrain the dynamics of the gas when a galaxy is resolved with only three resolution elements across its major axis and the SNR is larger than two \citep{3dbarolo}. Moreover, differently from the past, when a set of initial estimates for the galaxy main parameters (such as systemic velocity, position of the centre of a galaxy, position and inclination angles) was required to derive a rotation curve, state-of-the-art kinematic modelling software is able to perform a fit completely blind. 

For our study we use the latest fully automated version of $\rm^{3D}$Barolo (1.6.1), a tool for fitting 3D tilted-ring models to emission-line data-cubes \citep{3dbarolo}. 

\begin{figure*} 
\centering
   \includegraphics[scale=0.5]{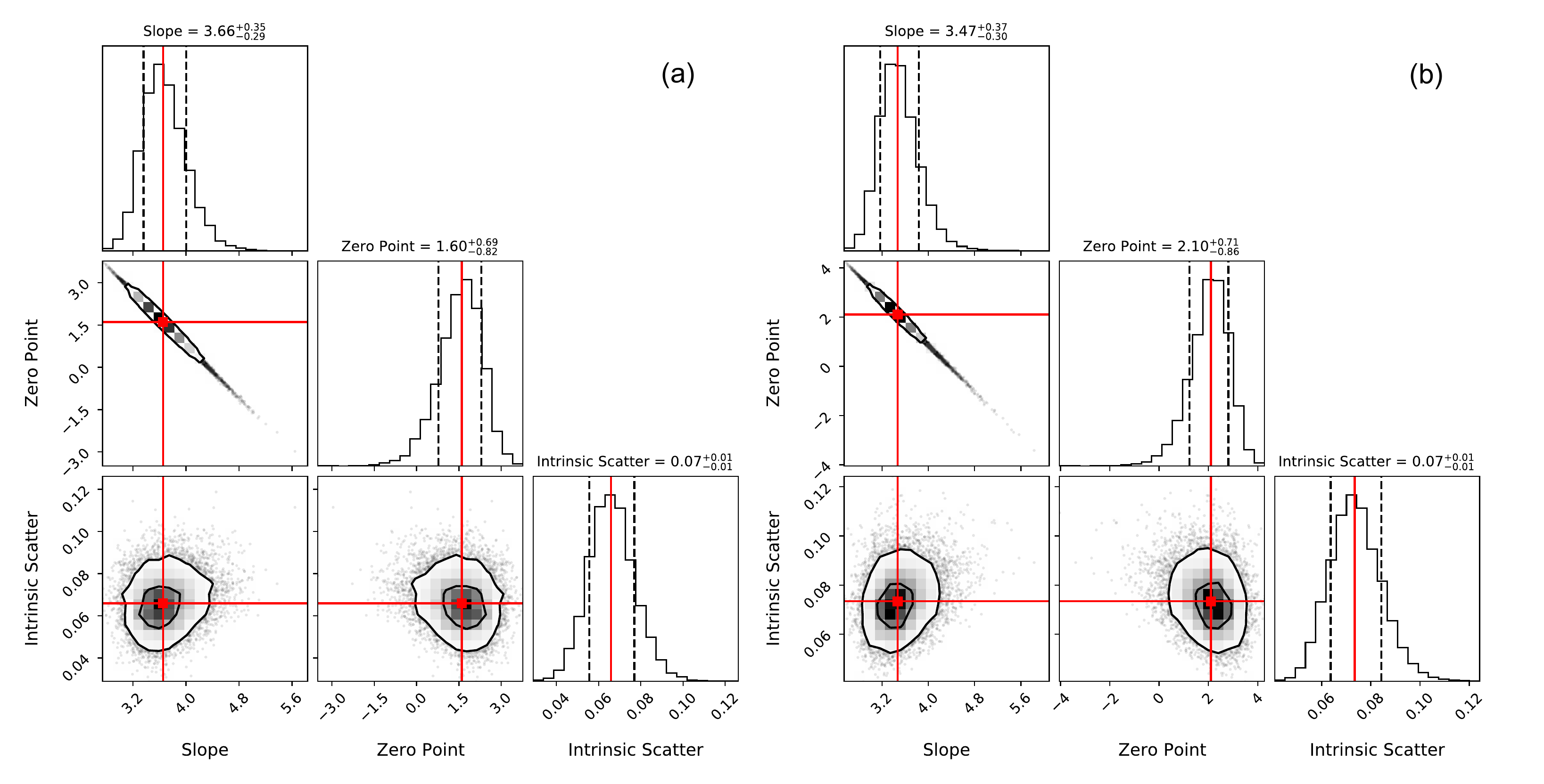}
   \caption{The posterior distributions of the slope, zero point and intrinsic scatter for the bTFR based on $\rm W_{50}$ (a) and $\rm V_{out}$ (b). Red crosses and solid lines show the maximum likelihood values. Black contours indicate 68 and 95 per cent confidence levels.}
\label{fig_corner}
\end{figure*}

\subsubsection{The sample and results}
\label{subsec:incl}
We construct our initial bTFr kinematic sample as follows. First, we use the tight relation between the \hi mass and \hi diameter \citep{wang2016} to define the radial extent of each galaxy\footnote{Our MIGHTEE-\hi galaxy sample has been shown to closely follow this relation during the data quality control check (Frank et al. in prep.).}, and retain those galaxies which have a size of at least with three resolution elements (see Table \ref{tbl_data} for the beam sizes). We then select galaxies with corrected $\rm W_{50} > 44$ \kms, since we are limited by the velocity resolution of the data. Finally, we select galaxies with $ i_{\hi}>20^{\circ}$, because geometric corrections to the rotational velocities are large and uncertain for the less inclined galaxies\footnote{ $i_{\hi}>20^{\circ}$ is somewhat more permissive than the classical approach where only galaxies with inclinations higher than 40$^{\circ}$ are included in the TFr samples  \citep{read2016}. However we do not find any systematic effects associated with including galaxies of lower inclinations.} \citep{mverheijen2001, ponomareva2016, lelli2019}. Consequently, our initial bTFr sample consists of 93 galaxies, which is $\sim 40\%$ of the entire sample of the MIGHTEE-\hi Early Science Data. 

We conduct the first completely blind run of $\rm^{3D}$Barolo on our initial sample, requesting two points of rotation curve per synthesised beam. In practice, the first run of $\rm^{3D}$Barolo finds a galaxy in the cubelet and estimates its radial extent, centre and systemic velocity. These parameters are then kept fixed, while rotation velocity, position and inclination angles, and velocity dispersion are fitted. 
While $\rm^{3D}$Barolo performs well in recovering various parameters of galaxies, including the position angle of each source, inclination angles have been proven to remain a challenge. Figure \ref{fig_incl3db} (a) shows the comparison between inclination angles obtained with $\rm^{3D}$Barolo's blind run and the \hi inclinations measured from \hi moment-0 maps (Section \ref{sec:incl}). While the majority of inclinations are within 10$^{\circ}$ of $i_{\hi}$, a clear trend is visible: $\rm^{3D}$Barolo tends to overestimate inclinations for low values of $i_{\hi}$ and underestimate them at higher inclinations. Following this, we perform a second run of the $\rm^{3D}$Barolo modelling, but this time we use the $i_{\hi}$ values as initial estimate and keep them unconstrained during the fit. In this case, $\rm^{3D}$Barolo recovers the initial inclination values within $\sim 5^{\circ}$ range, see Figure \ref{fig_incl3db} (b). Therefore, the inclination of a galaxy remains the main caveat for the blind kinematic modelling for low resolution data, and should be measured carefully in advance either from the \hi moment-0 maps, or from ancillary photometry. We note that the inclination measured from our optical ancillary data and the \hi data are consistent.
\begin{figure*} 
\centering
   \includegraphics[scale=0.7]{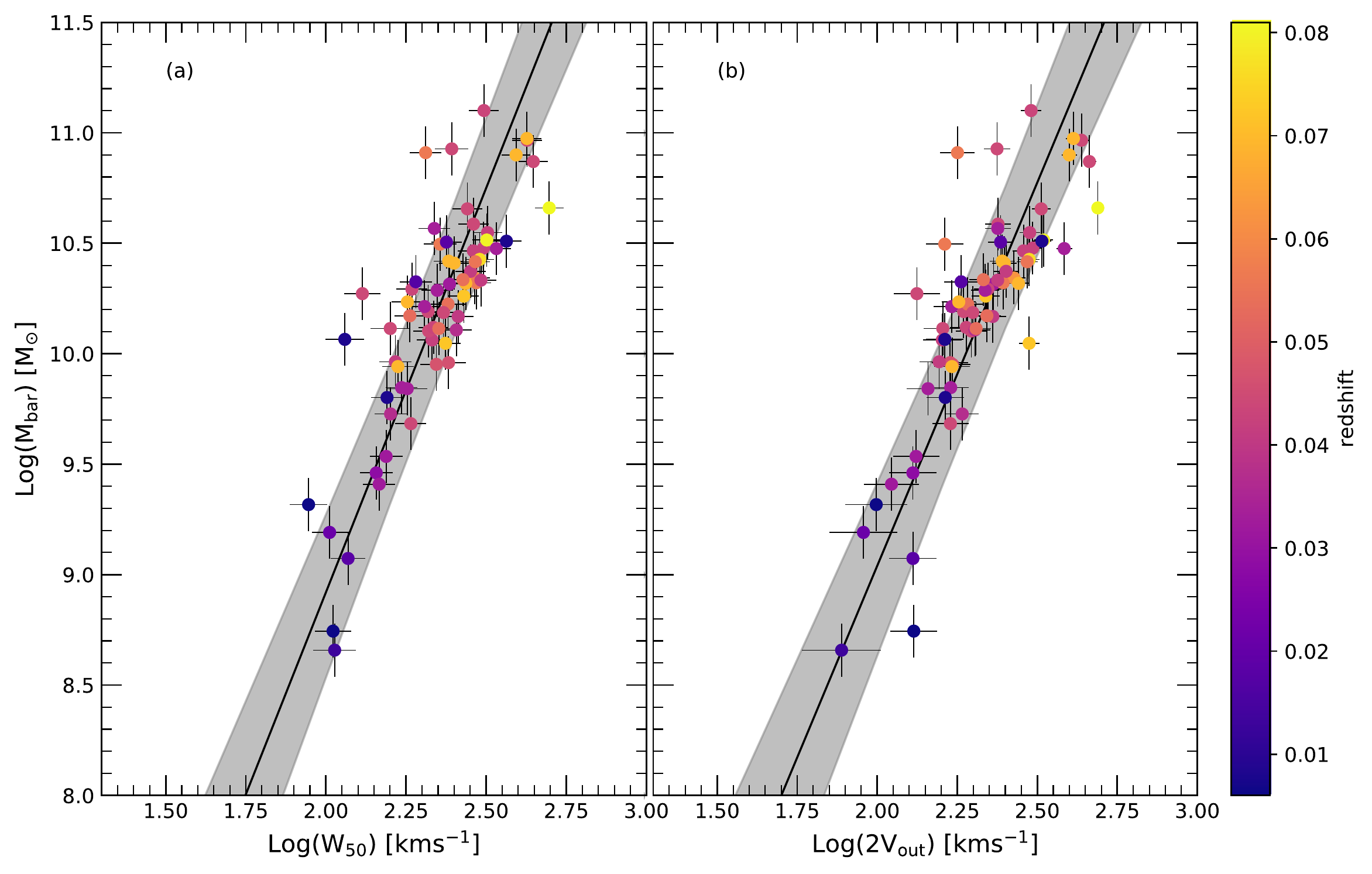}
   \caption{The bTFr with data points colour-coded as a function of redshift: panel (a) shows the bTFr based on the corrected  $\rm W_{50}$ as a rotational velocity measure; panel (b) shows the bTFr based on $\rm V_{out}$. The best fit is shown by a solid black line, and the 1-$\sigma$ uncertainty of the fit, sampled from the MCMC posteriors (Figure \ref{fig_corner}) is shown with the shaded area.}
\label{fig_tfrs}
\end{figure*}

We then perform an assessment of the resulting models by visually inspecting the data, model and residuals, as well as the resulting rotation curves projected on the position-velocity (p-v) diagrams (see Figures 5 and 6 in \citealt{maddox2021}). Only 67 out of 93 galaxies pass our visual assessment criteria\footnote{SNR and velocity resolution are the main reasons why $\rm ^{3D}$Barolo is not able to model the kinematics of some galaxies in our initial sample.}. They form the final bTFr sample which covers the $\rm 0.006 < z < 0.081$ redshift range\footnote{The machine-readable table, containing all measurements used in this analysis can be found at \href{https://bit.ly/3zKPwdW}{https://bit.ly/3zKPwdW}}. 

Resolved rotation curves represent the velocity of a galaxy as a function of radius. Normally, the velocities for the bTFr are measured either at the peak of the rotation curve ($\rm V_{max}$) or as an average velocity of its flat part ($\rm V_{flat}$). As stated previously, the use of the latter provides the tightest bTFR, since it is likely to better trace the DM halo potential \citep{ponomareva2018, lelli2019}. However, since our sample galaxies cover a wide range of radial extents ($\sim 30$ beams for large nearby spirals down to 3 beams for the sources at the highest redshift), we base our analysis on the rotational velocity measured at the outermost \hi radius $\rm V_{out} = V(R_{out})$ \citep{papastergis2016}. This is not only to simplify the analysis, but also to have a homogeneous measurement of the rotational velocity for all galaxies in the sample, given that the definition of the flat part of the rotation curve consisting of few points is very ambiguous \citep{lelli2016}. Therefore, our $\rm V_{out}$ measurements are a good representation of $\rm V_{flat}$ for objects that are well resolved, and of $\rm V_{max}$ for objects that are marginally resolved.
\begin{table} 
\centering
\begin{adjustbox}{width=0.45\textwidth}
\begin{tabular}{lcccc}   
\hline
\hline
Sample size: 67&$\rm W_{50}$&$\rm V_{out}$\\
 \hline
Slope& $3.66^{+0.35}_{-0.29}$ & $3.47^{+0.37}_{-0.30}$\Tstrut\\  
Zero Point& $1.60^{+0.69}_{-0.82}$ &  $2.10^{+0.71}_{-0.86}$&\Tstrut\\
Scatter ($\sigma$)& 0.33 & 0.32\\
Intrinsic Scatter ($\sigma_{\perp}$)& $0.07^{+0.01}_{-0.01}$ &  $0.07^{+0.01}_{-0.01}$\Bstrut\\
\hline
\end{tabular}
\end{adjustbox}
\caption{The statistical properties of the baryonic TFr based on two velocity measures, corrected $\rm W_{50}$ and $\rm V_{out}$, and obtained with the maximum likelihood orthogonal model.}
\label{tbl_scatter}
\end{table} 

The typical uncertainty on $\rm V_{out}$, including the uncertainty on inclination ($\sim 5^{\circ}$) and position angles ($\sim 7^{\circ}$), is consistent with a half channel width. Figure \ref{fig_W50-Vout} shows the comparison between the corrected $\rm W_{50}$ and the derived $\rm V_{out}$ for our final bTFr sample of galaxies. For a fair comparison of the two different rotational velocity measures we correct our $\rm W_{50}$ values by the kinematic inclinations produced during the second run of $\rm^{3D}$Barolo. We also propagate the uncertainty of the kinematic inclination angles into the uncertainty on $\rm W_{50}$. Overall the two measurements are in excellent agreement, with the mean of the difference ($\rm W_{50} - 2V_{out}$) equal to 0.02 dex, and the standard deviation equal to 0.06 dex.

\section{Results}
\label{sec:results}
In this section we present the statistical properties of the \hi-based baryonic TFr beyond $z=0$ for different rotational velocity measures ($\rm W_{50}$ \& $\rm V_{out}$).

\begin{figure*} 
\centering
   \includegraphics[scale=0.7]{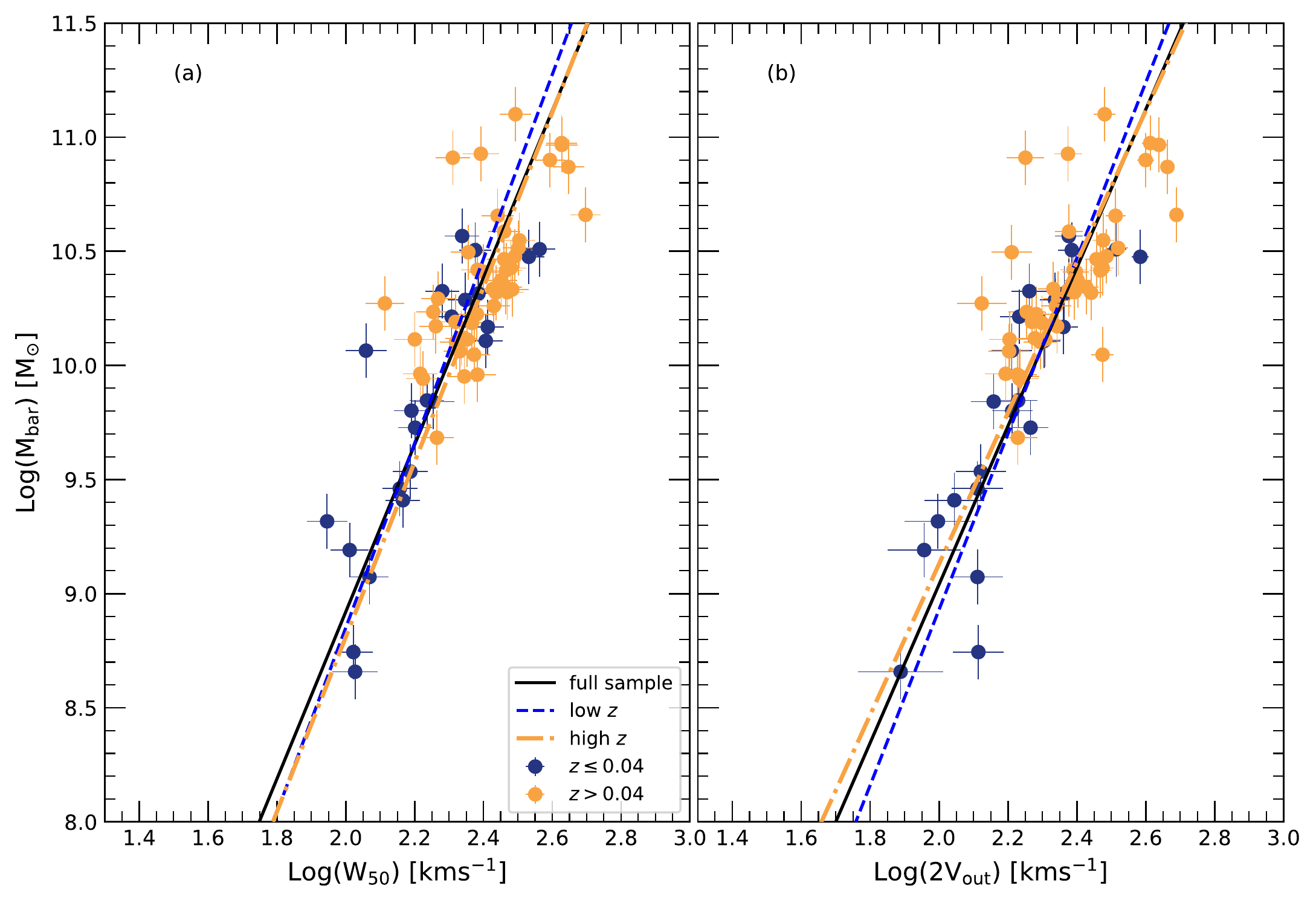}
   \caption{The bTFr based on corrected $\rm W_{50}$, panel (a) and $\rm V_{out}$, panel (b). The high-redshift galaxies ($z>0.04$) are shown with orange symbols, while the low-redshift galaxies ($z\leq0.04$) are shown with blue symbols. The best-fit for the full sample is shown with the straight black line, while the fits for low- and high-$z$ samples are shown with blue dashed and orange dashed-dotted lines respectively.}
\label{fig_malmq}
\end{figure*}

\subsection{Fitting method}
\label{sec:fits}
Even though the bTFr is a basic linear relation, the choice of the fit is not trivial because errors on both axes as well as intrinsic scatter should be taken into account. There have been a number of studies, suggesting that the choice of the fitting technique can significantly affect the bTFr \citep{bradford16, sorceguo16}. For example, it was shown that the slope of a TFr can be affected by Malmquist bias if a direct least squares fit is used \citep{TC12}, which can be resolved with the use of an inverse fit \citep{willick94}. Moreover, it is important to keep in mind that the vertical scatter of the bTFr is highly slope dependent. Hence, an intrinsically tight linear relation may have a large vertical scatter due to the steep slope \citep{mverheijen2001, ponomareva2017}. It is important to take all these effects into account when studying the statistical properties of the bTFr. \citet{lelli2019} compared three different types of linear fits and found that the maximum likelihood method which considered the orthogonal intrinsic scatter ($\sigma_{\perp}$) to be the preferred model. \citet{ponomareva2017} and \citet{stone2021} also found that the orthogonal linear regression model which minimises the orthogonal distances from the data points to the fitted line better describes the intrinsic properties of any TFr. 

For our study we perform linear fits with a maximum likelihood model that takes the errors in both directions into account and assumes a Gaussian distribution to describe the
intrinsic scatter along the perpendicular direction to the best-fit line. We follow the prescriptions described in \citet{lelli2019}, and use the standard affine-invariant ensemble sampler for Markov chain Monte Carlo (MCMC) {\it emcee} \citep{emcee} to map the posterior distributions of the main statistical properties: slope, zero point and intrinsic scatter.
For the fit we initialize the chains with 50 random walkers, run
1000 iterations and re-run the simulation with 1000 steps. The starting position of the walkers is set randomly within realistic ranges: slope [3.0, 5.0], zero point [1.0, 5.0] and intrinsic orthogonal scatter ($\sigma_{\perp}$) [0.01, 0.25]. The convergence of the chains is checked visually.

Figure \ref{fig_corner} shows these posterior distributions for the bTFr based on both $\rm W_{50}$, panel (a) and $\rm V_{out}$, panel (b). Table \ref{tbl_scatter} summarises the values obtained for the slope, zero point, $\sigma_{\perp}$ and standard observed vertical scatter ($\sigma$, see Eq. 3 in \citealt{ponomareva2017}). Asymmetric errors are estimated at the 68\% confidence level of the marginalized 1-D posterior distributions of the fitting parameters.

\subsection{The evolution of the bTFr at $ \rm 0 \leq z \leq 0.081$}

Figure \ref{fig_tfrs} shows the bTFr for our sample galaxies for the two velocity measures: $\rm W_{50}$, panel (a) and $\rm V_{out}$, panel (b) colour-coded as a function of redshift. It clearly shows that the majority of galaxies lie along the best-fit line, independent of the redshift. 
\begin{table} 
\centering
\begin{adjustbox}{width=0.45\textwidth}
\begin{tabular}{lcccc}   
\hline
\hline
Sample&\multicolumn{2}{c}{Slope}&\multicolumn{2}{c}{Zero Point}\\
& W$_{\rm 50}$&V$_{\rm out}$&W$_{\rm 50}$&V$_{\rm out}$\\
 \hline
$\rm 0 \leq z \leq 0.081$ (N=67)& $3.66^{+0.35}_{-0.29}$ &$3.47^{+0.37}_{-0.3}$& $1.60^{+0.69}_{-0.82}$& $2.10^{+0.71}_{-0.86}$\Tstrut\\  
$z\leq0.04$ (N=23)& $4.04^{+0.74}_{-0.52}$ &$3.85^{+0.76}_{-0.50}$& $0.77^{+1.16}_{-1.67}$&$1.23^{+1.15}_{-1.73}$\Tstrut\\
$z>0.04$ (N=44)& $3.83^{+0.84}_{-0.59}$ & $3.31^{+0.71}_{-0.49}$& $1.15^{+1.43}_{-2.03}$&$2.51^{+1.16}_{-1.70}$\Tstrut\Bstrut\\
\hline
\end{tabular}
\end{adjustbox}
\caption{The slope and zero point of the baryonic TFr based on $\rm W_{50}$ and $\rm V_{out}$ for the full, low-z and high-z samples, obtained with the maximum likelihood orthogonal model. N is the number of galaxies in the sample.}
\label{tbl_comfits}
\end{table}  

However, a slight Malmquist bias can be seen in our sample. The low redshift galaxies ($z\leq0.04$) lie within the 2\,dex mass range $\rm 8.6 \leq log(M_{bar})[M_{\odot}] \leq 10.6$, while galaxies at $z>0.04$ occupy a higher mass range from  $\rm 9.7 \leq log(M_{bar})[M_{\odot}] \leq 11.2$ (Figure \ref{fig_malmq}). To investigate if there is any evidence that the bTFr evolves over the last one billion years, we fit low- and high-$z$ samples separately in the same manner as described in Section \ref{sec:fits}. The resulting fits are shown in Figure \ref{fig_malmq} and Table \ref{tbl_comfits} for both $\rm W_{50}$ and $\rm V_{out}$. We find the slopes and zero points of both fits to be similar and consistent within the errors for $\rm W_{50}$, as well as consistent with the best-fit relation of the full sample. We therefore can conclude that our analysis does not suggest any evolution of the baryonic TFr during the last billion years. 

\subsection{The bTFr at $\rm 0 \leq z \leq 0.081$ and different velocity definitions}
Using the orthogonal maximum likelihood method we find the slopes of the relations based on the two different velocity measures to be consistent within the uncertainties (Figure \ref{fig_corner}, Table \ref{tbl_scatter}). However, it is somewhat surprising that $\rm W_{50}$ yields a slightly steeper slope than $\rm V_{out}$. Previous studies \citep{ponomareva2018, lelli2019}, which used different velocity definitions, find that the slope of the relation is the steepest when the flat part of the rotation curve ($\rm V_{flat}$) is used as the velocity measure, and becomes shallower when other definitions such as $\rm W_{50}$ or $\rm V_{max}$ are considered. These trends are well understood, and attributed to the shapes of the rotation curves \citep{mverheijen2001, noordermeer2007, ponomareva2017}. Maximum rotational velocities are measured in the inner parts of spiral galaxies and tend to overestimate the rotational velocity if compared to the outer regions where the rotation curves reach the flat part ($\rm V_{flat}$). For dwarf galaxies, rotation curves do not reach the flat part, and the maximum rotational velocities tend to underestimate $\rm V_{flat}$. Consequently, the slope of the relation will get shallower, because the velocity of high mass galaxies will be overestimated, shifting them to the right, while it will be underestimated for low mass galaxies, moving them to the left. 

Comparing our slopes and zero points to the \citet{lelli2019} relations based on $\rm W_{50}$, $\rm V_{flat}$ and $\rm V_{max}$, we find that our measurements for $\rm W_{50}$ are consistent within the errors, but our $\rm V_{out}$ results are more consistent with their relation based on $\rm V_{max}$ rather than $\rm V_{flat}$. This suggests that $\rm V_{out}$ in our study might be more representative of the maximum rotational velocity than of the rotational velocity measured at the flat part of the rotation curve. As mentioned previously, our 
$\rm V_{out}$ measurements are a mixture of $\rm V_{max}$ and $\rm V_{flat}$, depending on the spatial resolution of a galaxy. Figure \ref{fig_malmq} (b) shows that the slope for the low-z sample is steeper (although still consistent within the uncertainties), than for the high-z sample. Interestingly, the slope of the low-z sample is identical to the one found by \citealt{lelli2019} (3.85) with the relation based on $\rm V_{flat}$, while the high-z slope is more consistent with their slope for $\rm V_{max}$. We therefore conclude that our $\rm V_{out}$ is a good representation of $\rm V_{flat}$ for the low redshift galaxies, and mostly likely traces $\rm V_{max}$ for the high redshift galaxies which are naturally less resolved. This will be further tested when the new MIGHTEE survey data with higher velocity resolution becomes available.

Interestingly, we find no difference in the intrinsic scatter between the relations, based on different velocity measures as often found at $z=0$ (Table \ref{tbl_scatter}). We find both relations to be relatively tight $\sigma_{\perp}=0.07 \pm 0.01$, however not consistent with zero intrinsic scatter within 3$\sigma$ as suggested for $z=0$ \citep{lelli2019}. Future work with the full MIGHTEE-\hi sample and higher velocity resolution will provide further constraints on the intrinsic scatter of the bTFr beyond $z=0$.

In general, it is very encouraging that we find our results to be so consistent with the {\it SPARC} measurements. Not only because our \hi data cover different redshift range, and are of different spatial and velocity resolution, but also because of the different methods followed to estimate the stellar masses of galaxies. While {\it SPARC} uses a nearly constant mass-to-light ratio at 3.6 $\mu$m, we use the SED-fitting technique (Section \ref{stellar_mass}). It was shown by \citet{ponomareva2018} that the chosen method to estimate the stellar mass of spiral galaxies can have a significant effect on the statistical properties of the bTFr, with the SED-fitting producing masses which yield the highest intrinsic scatter. However, our result suggests that it is not the SED-fitting per se, but the quality of the photometric data that affect the statistical measurements. The high quality of the photometry in the MIGHTEE fields allows us to recover a small intrinsic scatter independently of the method used to estimate stellar mass.


\section{Summary and Conclusions}
\label{sec:summary}
In this paper we present the first study of the \hi-based baryonic Tully-Fisher relation over the last billion years using the Early Science data from the MIGHTEE survey. We are for the first time able to compare a higher redshift study to the local ($z=0$) bTFr, derived from one homogeneous single survey, where observations and analyses of sources were treated in the same manner. This study aims to investigate the statistical properties of the bTFr based on two different velocity measures: corrected $\rm W_{50}$ from the global \hi line profile, and $\rm V_{out}$ measured from the resolved \hi rotation curves. We derive resolved \hi rotation curves at $z > 0$ for the first time using a fully automated kinematic modelling software $\rm ^{3D}$Barolo. Our main results can be summarised as follows:

\begin{itemize}
\item The inclination of a galaxy is the main caveat for blind automated kinematic modelling, with $\rm ^{3D}$Barolo able to recover all galaxy parameters blindly, except for inclination. Therefore inclinations should be measured carefully in advance either from the \hi moment-0 maps or from the ancillary photometry, and provided as an initial estimate. 
\item We find no evidence for evolution of the bTFr over the last billion years, independent of the rotational velocity definition used ($\rm W_{50}$ or $\rm V_{out}$).

\item The bTFr over the last billion years is relatively tight with orthogonal intrinsic scatter $\sigma_{\perp}=0.07 \pm 0.01$, comparable to the {\it SPARC} sample at $z \simeq 0$ ($0.035\pm0.006$ for $\rm W_{50}$ and $0.04\pm0.006$ for $\rm V_{max}$), but not consistent with zero intrinsic scatter within 3$\sigma$. To be able to provide as accurate comparison as possible, we employ the same maximum likelihood fit as \citep{lelli2019}, that considers the orthogonal intrinsic scatter. 
\item We find the slopes of our relations to be consistent within the errors ($3.66^{+0.35}_{-0.29}$ for $\rm W_{50}$ and $3.47^{+0.37}_{-0.30}$ for $\rm V_{out}$) as well as consistent within the errors with the $z \simeq 0$ studies ($3.62 \pm 0.09$ for $\rm W_{50}$ and $3.52 \pm 0.07$ for $\rm V_{max}$). Unexpectedly, we find that the use of the corrected $\rm W_{50}$ results in a steeper slope, suggesting that our definition of $\rm V_{out}$ is consistent with $\rm V_{max}$ for high redshift galaxies and with $\rm V_{flat}$ for the low redshift sample. We will test this further when the full data with higher velocity resolution will be available from the complete MIGHTEE survey. Nevertheless, our results showing consistency with the largest study conducted at $z \simeq 0$ are a major breakthrough in studying the \hi-based bTFr and resolved \hi kinematics at higher redshifts. 
\end{itemize}

New observational facilities such as MeerKAT, have a potential to transform our knowledge about the \hi distribution and kinematics at high redshifts, well before the SKA era. The MIGHTEE Large Survey Program is well underway, and will give us an opportunity to extend the current study up to $z \simeq 0.5$, including the redshift evolution directly in the relation. Improved velocity resolution will be essential for the detailed study of not only the statistical properties of the bTFr, but also of the mass models of galaxies beyond $ \rm z = 0$. MIGHTEE is strongly complementary with the LADUMA survey \citep{maddox2016}, and combined they will provide crucial constraints on the evolution of \hi up to $z \sim1$, and to the current models of galaxy formation and evolution.

\section*{acknowledgements}
We thank the anonymous referee for valuable comments and suggestions which significantly improved this manuscript. We also thank Maarten Baes for providing useful comments.

The MeerKAT telescope is operated by the
South African Radio Astronomy Observatory, which is a facility of
the National Research Foundation, an agency of the Department of
Science and Innovation. We acknowledge use of the Inter-University
Institute for Data Intensive Astronomy (IDIA) data intensive research
cloud for data processing. IDIA is a South African university partnership
 involving the University of Cape Town, the University of Pretoria
and the University of the Western Cape. The authors acknowledge the
Centre for High Performance Computing (CHPC), South Africa, for
providing computational resources to this research project. This work
is based on data products from observations made with ESO Telescopes 
at the La Silla Paranal Observatory under ESO programme
ID 179.A-2005 (Ultra-VISTA) and ID 179.A- 2006 (VIDEO) and on
data products produced by CALET and the Cambridge Astronomy
Survey Unit on behalf of the Ultra-VISTA and VIDEO consortia.
Based on observations collected at the European Southern Observatory
under ESO programmes 179.A-2005 (UltraVISTA), and 179.A-2006 (VIDEO).
Based on observations obtained with MegaPrime/MegaCam, a joint
project of CFHT and CEA/IRFU, at the Canada-France-Hawaii Telescope 
(CFHT) which is operated by the National Research Council
(NRC) of Canada, the Institut National des Science de l’Univers of
the Centre National de la Recherche Scientifique (CNRS) of France,
and the University of Hawaii. This work is based in part on data products 
produced at Terapix available at the Canadian Astronomy Data
Centre as part of the Canada-France-Hawaii Telescope Legacy Survey,
 a collaborative project of NRC and CNRS.The Hyper SuprimeCam (HSC) 
 collaboration includes the astronomical communities of
Japan and Taiwan, and Princeton University. The HSC instrumentation and 
software were developed by the National Astronomical
Observatory of Japan (NAOJ), the Kavli Institute for the Physics
and Mathematics of the Universe (Kavli IPMU), the University of
Tokyo, the High Energy Accelerator Research Organization (KEK),
the Academia Sinica Institute for Astronomy and Astrophysics in Taiwan 
(ASIAA), and Princeton University. Funding was contributed by
the FIRST program from Japanese Cabinet Office, the Ministry of Education, 
Culture, Sports, Science and Technology (MEXT), the Japan
Society for the Promotion of Science (JSPS), Japan Science and Technology 
Agency (JST), the Toray Science Foundation, NAOJ, Kavli
IPMU, KEK, ASIAA, and Princeton University.

AAP and MJJ acknowledge support of the STFC consolidated grant ST/S000488/1. 
WM, RKK and SHAR are supported by the South African Research Chairs Initiative of the Department 
of Science and Technology and National Research Foundation.
MJJ and IH acknowledge support from the UK Science and Technology Facilities Council [ST/N000919/1].
MJJ, HP and IH acknowledge support from the South African Radio Astronomy Observatory (SARAO) which is a facility of the National Research Foundation (NRF), an agency of the Department of Science and Innovation.
MJJ and PWH acknowledge support from the Oxford Hintze Centre for Astrophysical 
Surveys which is funded through generous support from the Hintze Family Charitable Foundation. 
NM acknowledges support from the Bundesministerium f\"{u}r Bildung und Forschung (BMBF) award 05A20WM4.
RAAB acknowledges support from an STFC Ernest Rutherford Fellowship [grant number ST/T003596/1].
IP acknowledges financial support from the Italian Ministry of Foreign Affairs and International Cooperation (MAECI Grant Number ZA18GR02) and the South African Department of Science and Technology's National Research Foundation (DST-NRF Grant Number 113121) as part of the ISARP RADIOSKY2020 Joint Research Scheme.
EMDT was supported by the US National Science Foundation under grant 1616177.
EAKA is supported by the WISE research programme, which is financed by the Netherlands Organisation for Scientific Research (NWO).
NJA acknowledges funding from the Science and Technology Facilities Council (STFC) Grant Code ST/R505006/1

This research has made use of NASA’s Astrophysics Data System Bibliographic Services. This research
made use of Astropy\footnote{\href{http://www.astropy.org}{http://www.astropy.org}}.
a community-developed core Python package for Astronomy.

\section*{Data availability}
The machine-readable table, containing all measurements used in this analysis can be found \href{https://www.dropbox.com/s/b3tu76ecpa3g5br/master_sample_bTFr.txt?dl=0}{here}.
The complete fitting parameters produced for both bTFrs can be found \href{https://www.dropbox.com/s/540ay8zy9y5mx13/bTFRfit.zip?dl=0}{here}.

The MIGHTEE-\hi spectral cubes will be released as part of the first data release of the MIGHTEE survey, which will include cubelets of the sources discussed in this paper. The derived quantities from the multi-wavelength ancillary data will be released with the final data release of the VIDEO survey mid 2021. Alternative products are already available from the \emph{Herschel} Extragalactic Legacy Project (HELP; \citealt{shirley2021}) and also soon from the Deep Extragalactic VIsible Legacy Survey (DEVILS: \citealt{davies2021}).

\bibliographystyle{mnras}
\bibstyle{mnras}
\bibliography{MIGHTEE_TFR}

\begin{thebibliography}{}
\makeatletter
\relax
\def\mn@urlcharsother{\let\do\@makeother \do\$\do\&\do\#\do\^\do\_\do\%\do\~}
\def\mn@doi{\begingroup\mn@urlcharsother \@ifnextchar [ {\mn@doi@}
  {\mn@doi@[]}}
\def\mn@doi@[#1]#2{\def\@tempa{#1}\ifx\@tempa\@empty \href
  {http://dx.doi.org/#2} {doi:#2}\else \href {http://dx.doi.org/#2} {#1}\fi
  \endgroup}
\def\mn@eprint#1#2{\mn@eprint@#1:#2::\@nil}
\def\mn@eprint@arXiv#1{\href {http://arxiv.org/abs/#1} {{\tt arXiv:#1}}}
\def\mn@eprint@dblp#1{\href {http://dblp.uni-trier.de/rec/bibtex/#1.xml}
  {dblp:#1}}
\def\mn@eprint@#1:#2:#3:#4\@nil{\def\@tempa {#1}\def\@tempb {#2}\def\@tempc
  {#3}\ifx \@tempc \@empty \let \@tempc \@tempb \let \@tempb \@tempa \fi \ifx
  \@tempb \@empty \def\@tempb {arXiv}\fi \@ifundefined
  {mn@eprint@\@tempb}{\@tempb:\@tempc}{\expandafter \expandafter \csname
  mn@eprint@\@tempb\endcsname \expandafter{\@tempc}}}

\bibitem[\protect\citeauthoryear{Abril-Melgarejo et~al.,}{Abril-Melgarejo
  et~al.}{2021}]{Abril-Melgarejo2021}
Abril-Melgarejo V.,  et~al., 2021, The Tully-Fisher relation in dense groups at
  $z \sim 0.7$ in the MAGIC survey (\mn@eprint {arXiv} {2101.08069})

\bibitem[\protect\citeauthoryear{{Adams}, {Bowler}, {Jarvis}, {Hau{\ss}ler}  \&
  {Lagos}}{{Adams} et~al.}{2021}]{adams2021}
{Adams} N.~J.,  {Bowler} R.~A.~A.,  {Jarvis} M.~J.,  {Hau{\ss}ler} B.,
  {Lagos} C.~D.~P.,  2021, arXiv e-prints, \href
  {https://ui.adsabs.harvard.edu/abs/2021arXiv210107182A} {p. arXiv:2101.07182}

\bibitem[\protect\citeauthoryear{{Aihara} et~al.,}{{Aihara}
  et~al.}{2018}]{Aihara2018}
{Aihara} H.,  et~al., 2018, \mn@doi [\pasj] {10.1093/pasj/psx066}, \href
  {https://ui.adsabs.harvard.edu/abs/2018PASJ...70S...4A} {70, S4}

\bibitem[\protect\citeauthoryear{{Aihara} et~al.,}{{Aihara}
  et~al.}{2019}]{Aihara2019}
{Aihara} H.,  et~al., 2019, \mn@doi [\pasj] {10.1093/pasj/psz103}, \href
  {https://ui.adsabs.harvard.edu/abs/2019PASJ...71..114A} {71, 114}

\bibitem[\protect\citeauthoryear{{Arnett}}{{Arnett}}{1999}]{arnett1999}
{Arnett} D.,  1999, \mn@doi [\apss] {10.1023/A:1002131915162}, \href
  {https://ui.adsabs.harvard.edu/abs/1999Ap&SS.265...29A} {265, 29}

\bibitem[\protect\citeauthoryear{Arnouts, Cristiani, Moscardini, Matarrese,
  Lucchin, Fontana  \& Giallongo}{Arnouts et~al.}{1999}]{Arnouts1999}
Arnouts S.,  Cristiani S.,  Moscardini L.,  Matarrese S.,  Lucchin F.,  Fontana
  A.,   Giallongo E.,  1999, \mn@doi [MNRAS]
  {10.1046/j.1365-8711.1999.02978.x}, 310, 540–556

\bibitem[\protect\citeauthoryear{{Bacchini}, {Fraternali}, {Iorio}  \&
  {Pezzulli}}{{Bacchini} et~al.}{2019}]{Bacchini2019}
{Bacchini} C.,  {Fraternali} F.,  {Iorio} G.,   {Pezzulli} G.,  2019, \mn@doi
  [\aap] {10.1051/0004-6361/201834382}, \href
  {https://ui.adsabs.harvard.edu/abs/2019A&A...622A..64B} {622, A64}

\bibitem[\protect\citeauthoryear{{Blyth} et~al.,}{{Blyth}
  et~al.}{2016}]{blyth2016}
{Blyth} S.,  et~al., 2016, in MeerKAT Science: On the Pathway to the SKA. p.~4

\bibitem[\protect\citeauthoryear{{Bradford}, {Geha}  \& {van den
  Bosch}}{{Bradford} et~al.}{2016}]{bradford16}
{Bradford} J.~D.,  {Geha} M.~C.,   {van den Bosch} F.~C.,  2016, \mn@doi [\apj]
  {10.3847/0004-637X/832/1/11}, \href
  {http://adsabs.harvard.edu/abs/2016ApJ...832...11B} {832, 11}

\bibitem[\protect\citeauthoryear{{Buchner} et~al.,}{{Buchner}
  et~al.}{2014}]{Buchner2014}
{Buchner} J.,  et~al., 2014, \mn@doi [\aap] {10.1051/0004-6361/201322971},
  \href {https://ui.adsabs.harvard.edu/abs/2014A&A...564A.125B} {564, A125}

\bibitem[\protect\citeauthoryear{{Chung}, {van Gorkom}, {O'Neil}  \&
  {Bothun}}{{Chung} et~al.}{2002}]{Chung2002}
{Chung} A.,  {van Gorkom} J.~H.,  {O'Neil} K.,   {Bothun} G.~D.,  2002, \mn@doi
  [\aj] {10.1086/339979}, \href
  {https://ui.adsabs.harvard.edu/abs/2002AJ....123.2387C} {123, 2387}

\bibitem[\protect\citeauthoryear{Comrie et~al.,}{Comrie et~al.}{2020}]{carta}
Comrie A.,  et~al., 2020, {CARTA: The Cube Analysis and Rendering Tool for
  Astronomy}, \mn@doi{10.5281/zenodo.4034416}, \url
  {https://doi.org/10.5281/zenodo.4034416}

\bibitem[\protect\citeauthoryear{{Courteau}, {Andersen}, {Bershady},
  {MacArthur}  \& {Rix}}{{Courteau} et~al.}{2003}]{courteau03}
{Courteau} S.,  {Andersen} D.~R.,  {Bershady} M.~A.,  {MacArthur} L.~A.,
  {Rix} H.-W.,  2003, \mn@doi [\apj] {10.1086/376754}, \href
  {https://ui.adsabs.harvard.edu/abs/2003ApJ...594..208C} {594, 208}

\bibitem[\protect\citeauthoryear{{Courtois}, {Hoffman}, {Tully}  \&
  {Gottl{\"o}ber}}{{Courtois} et~al.}{2012}]{courtois2012}
{Courtois} H.~M.,  {Hoffman} Y.,  {Tully} R.~B.,   {Gottl{\"o}ber} S.,  2012,
  \mn@doi [\apj] {10.1088/0004-637X/744/1/43}, \href
  {https://ui.adsabs.harvard.edu/abs/2012ApJ...744...43C} {744, 43}

\bibitem[\protect\citeauthoryear{{Cuillandre} et~al.,}{{Cuillandre}
  et~al.}{2012}]{Cuillandre2012}
{Cuillandre} J.-C.~J.,  et~al., 2012, in {Peck} A.~B.,  {Seaman} R.~L.,
  {Comeron} F.,  eds,  Society of Photo-Optical Instrumentation Engineers
  (SPIE) Conference Series Vol. 8448, Observatory Operations: Strategies,
  Processes, and Systems IV. p. 84480M, \mn@doi{10.1117/12.925584}

\bibitem[\protect\citeauthoryear{{Dav{\'e}}, {Angl{\'e}s-Alc{\'a}zar},
  {Narayanan}, {Li}, {Rafieferantsoa}  \& {Appleby}}{{Dav{\'e}}
  et~al.}{2019}]{dave2019}
{Dav{\'e}} R.,  {Angl{\'e}s-Alc{\'a}zar} D.,  {Narayanan} D.,  {Li} Q.,
  {Rafieferantsoa} M.~H.,   {Appleby} S.,  2019, \mn@doi [\mnras]
  {10.1093/mnras/stz937}, \href
  {https://ui.adsabs.harvard.edu/abs/2019MNRAS.486.2827D} {486, 2827}

\bibitem[\protect\citeauthoryear{{Davies} et~al.,}{{Davies}
  et~al.}{2021}]{davies2021}
{Davies} L.~J.~M.,  et~al., 2021, arXiv e-prints, \href
  {https://ui.adsabs.harvard.edu/abs/2021arXiv210606241D} {p. arXiv:2106.06241}

\bibitem[\protect\citeauthoryear{{Desmond}}{{Desmond}}{2012}]{desmond12}
{Desmond} H.,  2012, preprint, \href
  {http://adsabs.harvard.edu/abs/2012arXiv1204.1497D} {} (\mn@eprint {arXiv}
  {1204.1497})

\bibitem[\protect\citeauthoryear{{Desmond} \& {Wechsler}}{{Desmond} \&
  {Wechsler}}{2015}]{desmond2015}
{Desmond} H.,  {Wechsler} R.~H.,  2015, \mn@doi [\mnras]
  {10.1093/mnras/stv1978}, \href
  {https://ui.adsabs.harvard.edu/abs/2015MNRAS.454..322D} {454, 322}

\bibitem[\protect\citeauthoryear{{Di Teodoro} \& {Fraternali}}{{Di Teodoro} \&
  {Fraternali}}{2015}]{3dbarolo}
{Di Teodoro} E.~M.,  {Fraternali} F.,  2015, \mn@doi [\mnras]
  {10.1093/mnras/stv1213}, \href
  {https://ui.adsabs.harvard.edu/abs/2015MNRAS.451.3021D} {451, 3021}

\bibitem[\protect\citeauthoryear{{Dupuy}, {Courtois}, {Guinet}, {Tully}  \&
  {Kourkchi}}{{Dupuy} et~al.}{2021}]{cosmic4}
{Dupuy} A.,  {Courtois} H.~M.,  {Guinet} D.,  {Tully} R.~B.,   {Kourkchi} E.,
  2021, \mn@doi [\aap] {10.1051/0004-6361/202039025}, \href
  {https://ui.adsabs.harvard.edu/abs/2021A&A...646A.113D} {646, A113}

\bibitem[\protect\citeauthoryear{{Dutton}}{{Dutton}}{2012}]{dutton2012}
{Dutton} A.~A.,  2012, \mn@doi [\mnras] {10.1111/j.1365-2966.2012.21469.x},
  \href {https://ui.adsabs.harvard.edu/abs/2012MNRAS.424.3123D} {424, 3123}

\bibitem[\protect\citeauthoryear{{Feroz} \& {Hobson}}{{Feroz} \&
  {Hobson}}{2008}]{Feroz2008}
{Feroz} F.,  {Hobson} M.~P.,  2008, \mn@doi [\mnras]
  {10.1111/j.1365-2966.2007.12353.x}, \href
  {https://ui.adsabs.harvard.edu/abs/2008MNRAS.384..449F} {384, 449}

\bibitem[\protect\citeauthoryear{{Feroz}, {Hobson}  \& {Bridges}}{{Feroz}
  et~al.}{2009}]{Feroz2009}
{Feroz} F.,  {Hobson} M.~P.,   {Bridges} M.,  2009, \mn@doi [\mnras]
  {10.1111/j.1365-2966.2009.14548.x}, \href
  {https://ui.adsabs.harvard.edu/abs/2009MNRAS.398.1601F} {398, 1601}

\bibitem[\protect\citeauthoryear{{Foreman-Mackey}, {Hogg}, {Lang}  \&
  {Goodman}}{{Foreman-Mackey} et~al.}{2013}]{emcee}
{Foreman-Mackey} D.,  {Hogg} D.~W.,  {Lang} D.,   {Goodman} J.,  2013, \mn@doi
  [\pasp] {10.1086/670067}, \href
  {https://ui.adsabs.harvard.edu/abs/2013PASP..125..306F} {125, 306}

\bibitem[\protect\citeauthoryear{{Frank}, {de Blok}, {Walter}, {Leroy}  \&
  {Carignan}}{{Frank} et~al.}{2016}]{frank2016}
{Frank} B.~S.,  {de Blok} W.~J.~G.,  {Walter} F.,  {Leroy} A.,   {Carignan} C.,
   2016, \mn@doi [\aj] {10.3847/0004-6256/151/4/94}, \href
  {https://ui.adsabs.harvard.edu/abs/2016AJ....151...94F} {151, 94}

\bibitem[\protect\citeauthoryear{{Giovanelli}, {Haynes}, {da Costa},
  {Freudling}, {Salzer}  \& {Wegner}}{{Giovanelli}
  et~al.}{1997}]{giovanelli1997}
{Giovanelli} R.,  {Haynes} M.~P.,  {da Costa} L.~N.,  {Freudling} W.,  {Salzer}
  J.~J.,   {Wegner} G.,  1997, \mn@doi [\apjl] {10.1086/310521}, \href
  {https://ui.adsabs.harvard.edu/abs/1997ApJ...477L...1G} {477, L1}

\bibitem[\protect\citeauthoryear{{Glowacki}, {Elson}  \& {Dav{\'e}}}{{Glowacki}
  et~al.}{2020a}]{Glowacki2020b}
{Glowacki} M.,  {Elson} E.,   {Dav{\'e}} R.,  2020a, arXiv e-prints, \href
  {https://ui.adsabs.harvard.edu/abs/2020arXiv201108866G} {p. arXiv:2011.08866}

\bibitem[\protect\citeauthoryear{{Glowacki}, {Elson}  \& {Dav{\'e}}}{{Glowacki}
  et~al.}{2020b}]{Glowacki2020a}
{Glowacki} M.,  {Elson} E.,   {Dav{\'e}} R.,  2020b, \mn@doi [\mnras]
  {10.1093/mnras/staa2616}, \href
  {https://ui.adsabs.harvard.edu/abs/2020MNRAS.498.3687G} {498, 3687}

\bibitem[\protect\citeauthoryear{{Gogate}, {Verheijen}, {Deshev}, {van Gorkom},
  {Montero-Casta{\~n}o}, {van der Hulst}, {Jaff{\'e}}  \& {Poggianti}}{{Gogate}
  et~al.}{2020}]{budhies2020}
{Gogate} A.~R.,  {Verheijen} M.~A.~W.,  {Deshev} B.~Z.,  {van Gorkom} J.~H.,
  {Montero-Casta{\~n}o} M.,  {van der Hulst} J.~M.,  {Jaff{\'e}} Y.~L.,
  {Poggianti} B.~M.,  2020, \mn@doi [\mnras] {10.1093/mnras/staa1680}, \href
  {https://ui.adsabs.harvard.edu/abs/2020MNRAS.496.3531G} {496, 3531}

\bibitem[\protect\citeauthoryear{{Hess} et~al.,}{{Hess}
  et~al.}{2019}]{chiles2019}
{Hess} K.~M.,  et~al., 2019, \mn@doi [\mnras] {10.1093/mnras/sty3421}, \href
  {https://ui.adsabs.harvard.edu/abs/2019MNRAS.484.2234H} {484, 2234}

\bibitem[\protect\citeauthoryear{Ilbert et~al.,}{Ilbert
  et~al.}{2006}]{Ilbert2006}
Ilbert O.,  et~al., 2006, \mn@doi [A\&A] {10.1051/0004-6361:20065138}, 457,
  841–856

\bibitem[\protect\citeauthoryear{{Iorio}, {Fraternali}, {Nipoti}, {Di Teodoro},
  {Read}  \& {Battaglia}}{{Iorio} et~al.}{2017}]{iorio2017}
{Iorio} G.,  {Fraternali} F.,  {Nipoti} C.,  {Di Teodoro} E.,  {Read} J.~I.,
  {Battaglia} G.,  2017, \mn@doi [\mnras] {10.1093/mnras/stw3285}, \href
  {https://ui.adsabs.harvard.edu/abs/2017MNRAS.466.4159I} {466, 4159}

\bibitem[\protect\citeauthoryear{{Jarvis} et~al.,}{{Jarvis}
  et~al.}{2013}]{jarvis2013}
{Jarvis} M.~J.,  et~al., 2013, \mn@doi [\mnras] {10.1093/mnras/sts118}, \href
  {https://ui.adsabs.harvard.edu/abs/2013MNRAS.428.1281J} {428, 1281}

\bibitem[\protect\citeauthoryear{{Jarvis} et~al.,}{{Jarvis}
  et~al.}{2016}]{jarvis2016}
{Jarvis} M.,  et~al., 2016, in MeerKAT Science: On the Pathway to the SKA. p.~6
  (\mn@eprint {arXiv} {1709.01901})

\bibitem[\protect\citeauthoryear{{Jonas}}{{Jonas}}{2009}]{Jonas_2009}
{Jonas} J.~L.,  2009, \mn@doi [IEEE Proceedings] {10.1109/JPROC.2009.2020713},
  \href {https://ui.adsabs.harvard.edu/abs/2009IEEEP..97.1522J} {97, 1522}

\bibitem[\protect\citeauthoryear{{J{\'o}zsa}, {Kenn}, {Klein}  \&
  {Oosterloo}}{{J{\'o}zsa} et~al.}{2007}]{tirific}
{J{\'o}zsa} G.~I.~G.,  {Kenn} F.,  {Klein} U.,   {Oosterloo} T.~A.,  2007,
  \mn@doi [\aap] {10.1051/0004-6361:20066164}, \href
  {https://ui.adsabs.harvard.edu/abs/2007A&A...468..731J} {468, 731}

\bibitem[\protect\citeauthoryear{{Karachentsev}, {Kaisina}  \& {Kashibadze
  Nasonova}}{{Karachentsev} et~al.}{2017}]{Karachentsev2016}
{Karachentsev} I.~D.,  {Kaisina} E.~I.,   {Kashibadze Nasonova} O.~G.,  2017,
  \mn@doi [\aj] {10.3847/1538-3881/153/1/6}, \href
  {https://ui.adsabs.harvard.edu/abs/2017AJ....153....6K} {153, 6}

\bibitem[\protect\citeauthoryear{{Lelli}, {McGaugh}  \& {Schombert}}{{Lelli}
  et~al.}{2016a}]{lelli2016b}
{Lelli} F.,  {McGaugh} S.~S.,   {Schombert} J.~M.,  2016a, \mn@doi [\aj]
  {10.3847/0004-6256/152/6/157}, \href
  {https://ui.adsabs.harvard.edu/abs/2016AJ....152..157L} {152, 157}

\bibitem[\protect\citeauthoryear{{Lelli}, {McGaugh}  \& {Schombert}}{{Lelli}
  et~al.}{2016b}]{lelli2016}
{Lelli} F.,  {McGaugh} S.~S.,   {Schombert} J.~M.,  2016b, \mn@doi [\apjl]
  {10.3847/2041-8205/816/1/L14}, \href
  {http://adsabs.harvard.edu/abs/2016ApJ...816L..14L} {816, L14}

\bibitem[\protect\citeauthoryear{{Lelli}, {McGaugh}, {Schombert}, {Desmond}  \&
  {Katz}}{{Lelli} et~al.}{2019}]{lelli2019}
{Lelli} F.,  {McGaugh} S.~S.,  {Schombert} J.~M.,  {Desmond} H.,   {Katz} H.,
  2019, \mn@doi [\mnras] {10.1093/mnras/stz205}, \href
  {https://ui.adsabs.harvard.edu/abs/2019MNRAS.484.3267L} {484, 3267}

\bibitem[\protect\citeauthoryear{{Maddox}, {Jarvis}  \& {Oosterloo}}{{Maddox}
  et~al.}{2016}]{maddox2016}
{Maddox} N.,  {Jarvis} M.~J.,   {Oosterloo} T.~A.,  2016, \mn@doi [\mnras]
  {10.1093/mnras/stw1164}, \href
  {https://ui.adsabs.harvard.edu/abs/2016MNRAS.460.3419M} {460, 3419}

\bibitem[\protect\citeauthoryear{{Maddox} et~al.,}{{Maddox}
  et~al.}{2021}]{maddox2021}
{Maddox} N.,  et~al., 2021, \mn@doi [\aap] {10.1051/0004-6361/202039655}, \href
  {https://ui.adsabs.harvard.edu/abs/2021A&A...646A..35M} {646, A35}

\bibitem[\protect\citeauthoryear{{Mancera Pi{\~n}a} et~al.,}{{Mancera Pi{\~n}a}
  et~al.}{2019}]{pavel2019}
{Mancera Pi{\~n}a} P.~E.,  et~al., 2019, \mn@doi [\apjl]
  {10.3847/2041-8213/ab40c7}, \href
  {https://ui.adsabs.harvard.edu/abs/2019ApJ...883L..33M} {883, L33}

\bibitem[\protect\citeauthoryear{{Mancera Pi{\~n}a} et~al.,}{{Mancera Pi{\~n}a}
  et~al.}{2020}]{pavel2020}
{Mancera Pi{\~n}a} P.~E.,  et~al., 2020, \mn@doi [\mnras]
  {10.1093/mnras/staa1256}, \href
  {https://ui.adsabs.harvard.edu/abs/2020MNRAS.495.3636M} {495, 3636}

\bibitem[\protect\citeauthoryear{{McCracken} et~al.,}{{McCracken}
  et~al.}{2012}]{McCracken2012}
{McCracken} H.~J.,  et~al., 2012, \mn@doi [\aap] {10.1051/0004-6361/201219507},
  \href {https://ui.adsabs.harvard.edu/abs/2012A&A...544A.156M} {544, A156}

\bibitem[\protect\citeauthoryear{{McGaugh}}{{McGaugh}}{2012}]{mcgaugh12}
{McGaugh} S.~S.,  2012, \mn@doi [\aj] {10.1088/0004-6256/143/2/40}, \href
  {http://adsabs.harvard.edu/abs/2012AJ....143...40M} {143, 40}

\bibitem[\protect\citeauthoryear{{McGaugh}, {Schombert}, {Bothun}  \& {de
  Blok}}{{McGaugh} et~al.}{2000}]{mcgaugh00}
{McGaugh} S.~S.,  {Schombert} J.~M.,  {Bothun} G.~D.,   {de Blok} W.~J.~G.,
  2000, \mn@doi [\apjl] {10.1086/312628}, \href
  {http://adsabs.harvard.edu/abs/2000ApJ...533L..99M} {533, L99}

\bibitem[\protect\citeauthoryear{{McMullin}, {Waters}, {Schiebel}, {Young}  \&
  {Golap}}{{McMullin} et~al.}{2007}]{McMullin2007}
{McMullin} J.~P.,  {Waters} B.,  {Schiebel} D.,  {Young} W.,   {Golap} K.,
  2007, in {Shaw} R.~A.,  {Hill} F.,   {Bell} D.~J.,  eds,  Astronomical
  Society of the Pacific Conference Series Vol. 376, Astronomical Data Analysis
  Software and Systems XVI. p.~127

\bibitem[\protect\citeauthoryear{{Meyer}}{{Meyer}}{2009}]{meyer2009}
{Meyer} M.,  2009, in Panoramic Radio Astronomy: Wide-field 1-2 GHz Research on
  Galaxy Evolution. p.~15 (\mn@eprint {arXiv} {0912.2167})

\bibitem[\protect\citeauthoryear{Meyer, Robotham, Obreschkow, Westmeier, Duffy
  \& Staveley-Smith}{Meyer et~al.}{2017}]{meyer2017}
Meyer M.,  Robotham A.,  Obreschkow D.,  Westmeier T.,  Duffy A.~R.,
  Staveley-Smith L.,  2017, \mn@doi [Publications of the Astronomical Society
  of Australia] {10.1017/pasa.2017.31}, 34

\bibitem[\protect\citeauthoryear{{Mihalas} \& {Binney}}{{Mihalas} \&
  {Binney}}{1981}]{mihalas1981}
{Mihalas} D.,  {Binney} J.,  1981, Science, \href
  {https://ui.adsabs.harvard.edu/abs/1981Sci...214..829M} {214, 829}

\bibitem[\protect\citeauthoryear{{Noordermeer}, {van der Hulst}, {Sancisi},
  {Swaters}  \& {van Albada}}{{Noordermeer} et~al.}{2007}]{noordermeer2007}
{Noordermeer} E.,  {van der Hulst} J.~M.,  {Sancisi} R.,  {Swaters} R.~S.,
  {van Albada} T.~S.,  2007, \mn@doi [\mnras]
  {10.1111/j.1365-2966.2007.11533.x}, \href
  {https://ui.adsabs.harvard.edu/abs/2007MNRAS.376.1513N} {376, 1513}

\bibitem[\protect\citeauthoryear{{Obreschkow} \& {Meyer}}{{Obreschkow} \&
  {Meyer}}{2013}]{obreschkow2013}
{Obreschkow} D.,  {Meyer} M.,  2013, \mn@doi [\apj]
  {10.1088/0004-637X/777/2/140}, \href
  {https://ui.adsabs.harvard.edu/abs/2013ApJ...777..140O} {777, 140}

\bibitem[\protect\citeauthoryear{{Pan}, {Jarvis}, {Allison}, {Heywood},
  {Santos}, {Maddox}, {Frank}  \& {Kang}}{{Pan} et~al.}{2020}]{Pan2020}
{Pan} H.,  {Jarvis} M.~J.,  {Allison} J.~R.,  {Heywood} I.,  {Santos} M.~G.,
  {Maddox} N.,  {Frank} B.~S.,   {Kang} X.,  2020, \mn@doi [\mnras]
  {10.1093/mnras/stz3030}, \href
  {https://ui.adsabs.harvard.edu/abs/2020MNRAS.491.1227P} {491, 1227}

\bibitem[\protect\citeauthoryear{{Pan}, {Jarvis}, {Ponomareva}, {Santos},
  {Allison}, {Maddox}  \& {Frank}}{{Pan} et~al.}{2021}]{Pan2021}
{Pan} H.,  {Jarvis} M.~J.,  {Ponomareva} A.~A.,  {Santos} M.~G.,  {Allison}
  J.~R.,  {Maddox} N.,   {Frank} B.~S.,  2021, arXiv e-prints, \href
  {https://ui.adsabs.harvard.edu/abs/2021arXiv210904273P} {p. arXiv:2109.04273}

\bibitem[\protect\citeauthoryear{{Papastergis} \& {Shankar}}{{Papastergis} \&
  {Shankar}}{2016}]{papastergis2016}
{Papastergis} E.,  {Shankar} F.,  2016, \mn@doi [\aap]
  {10.1051/0004-6361/201527854}, \href
  {https://ui.adsabs.harvard.edu/abs/2016A&A...591A..58P} {591, A58}

\bibitem[\protect\citeauthoryear{{Ponomareva}, {Verheijen}  \&
  {Bosma}}{{Ponomareva} et~al.}{2016}]{ponomareva2016}
{Ponomareva} A.~A.,  {Verheijen} M. A.~W.,   {Bosma} A.,  2016, \mn@doi
  [\mnras] {10.1093/mnras/stw2213}, \href
  {https://ui.adsabs.harvard.edu/abs/2016MNRAS.463.4052P} {463, 4052}

\bibitem[\protect\citeauthoryear{{Ponomareva}, {Verheijen}, {Peletier}  \&
  {Bosma}}{{Ponomareva} et~al.}{2017}]{ponomareva2017}
{Ponomareva} A.~A.,  {Verheijen} M. A.~W.,  {Peletier} R.~F.,   {Bosma} A.,
  2017, \mn@doi [\mnras] {10.1093/mnras/stx1018}, \href
  {https://ui.adsabs.harvard.edu/abs/2017MNRAS.469.2387P} {469, 2387}

\bibitem[\protect\citeauthoryear{{Ponomareva}, {Verheijen}, {Papastergis},
  {Bosma}  \& {Peletier}}{{Ponomareva} et~al.}{2018}]{ponomareva2018}
{Ponomareva} A.~A.,  {Verheijen} M. A.~W.,  {Papastergis} E.,  {Bosma} A.,
  {Peletier} R.~F.,  2018, \mn@doi [\mnras] {10.1093/mnras/stx3066}, \href
  {https://ui.adsabs.harvard.edu/abs/2018MNRAS.474.4366P} {474, 4366}

\bibitem[\protect\citeauthoryear{{Ramatsoku} et~al.,}{{Ramatsoku}
  et~al.}{2016}]{mpati2016}
{Ramatsoku} M.,  et~al., 2016, \mn@doi [\mnras] {10.1093/mnras/stw968}, \href
  {https://ui.adsabs.harvard.edu/abs/2016MNRAS.460..923R} {460, 923}

\bibitem[\protect\citeauthoryear{{Ranchod} et~al.,}{{Ranchod}
  et~al.}{2021}]{ranchod2021}
{Ranchod} S.,  et~al., 2021, \mn@doi [\mnras] {10.1093/mnras/stab1817}, \href
  {https://ui.adsabs.harvard.edu/abs/2021MNRAS.506.2753R} {506, 2753}

\bibitem[\protect\citeauthoryear{{Read}, {Iorio}, {Agertz}  \&
  {Fraternali}}{{Read} et~al.}{2016}]{read2016}
{Read} J.~I.,  {Iorio} G.,  {Agertz} O.,   {Fraternali} F.,  2016, \mn@doi
  [\mnras] {10.1093/mnras/stw1876}, \href
  {https://ui.adsabs.harvard.edu/abs/2016MNRAS.462.3628R} {462, 3628}

\bibitem[\protect\citeauthoryear{{Robitaille}, {Ginsburg}, {Beaumont}, {Leroy}
  \& {Rosolowsky}}{{Robitaille} et~al.}{2016}]{speccube}
{Robitaille} T.,  {Ginsburg} A.,  {Beaumont} C.,  {Leroy} A.,   {Rosolowsky}
  E.,  2016, {spectral-cube: Read and analyze astrophysical spectral data
  cubes} (\mn@eprint {ascl} {1609.017})

\bibitem[\protect\citeauthoryear{{Rogstad}, {Lockhart}  \& {Wright}}{{Rogstad}
  et~al.}{1974}]{rogstad1974}
{Rogstad} D.~H.,  {Lockhart} I.~A.,   {Wright} M.~C.~H.,  1974, \mn@doi [\apj]
  {10.1086/153164}, \href
  {https://ui.adsabs.harvard.edu/abs/1974ApJ...193..309R} {193, 309}

\bibitem[\protect\citeauthoryear{{Shirley} et~al.,}{{Shirley}
  et~al.}{2021}]{shirley2021}
{Shirley} R.,  et~al., 2021, arXiv e-prints, \href
  {https://ui.adsabs.harvard.edu/abs/2021arXiv210505659S} {p. arXiv:2105.05659}

\bibitem[\protect\citeauthoryear{{Sorce} \& {Guo}}{{Sorce} \&
  {Guo}}{2016}]{sorceguo16}
{Sorce} J.~G.,  {Guo} Q.,  2016, \mn@doi [\mnras] {10.1093/mnras/stw341}, \href
  {http://adsabs.harvard.edu/abs/2016MNRAS.458.2667S} {458, 2667}

\bibitem[\protect\citeauthoryear{{Sorce} et~al.,}{{Sorce}
  et~al.}{2013}]{sorce13}
{Sorce} J.~G.,  et~al., 2013, \mn@doi [\apj] {10.1088/0004-637X/765/2/94},
  \href {http://adsabs.harvard.edu/abs/2013ApJ...765...94S} {765, 94}

\bibitem[\protect\citeauthoryear{{Stone}, {Courteau}  \& {Arora}}{{Stone}
  et~al.}{2021}]{stone2021}
{Stone} C.,  {Courteau} S.,   {Arora} N.,  2021, \mn@doi [\apj]
  {10.3847/1538-4357/abebe4}, \href
  {https://ui.adsabs.harvard.edu/abs/2021ApJ...912...41S} {912, 41}

\bibitem[\protect\citeauthoryear{{Tiley} et~al.,}{{Tiley}
  et~al.}{2016}]{tiley2016}
{Tiley} A.~L.,  et~al., 2016, \mn@doi [\mnras] {10.1093/mnras/stw936}, \href
  {https://ui.adsabs.harvard.edu/abs/2016MNRAS.460..103T} {460, 103}

\bibitem[\protect\citeauthoryear{{Tiley} et~al.,}{{Tiley}
  et~al.}{2019}]{tiley19}
{Tiley} A.~L.,  et~al., 2019, \mn@doi [\mnras] {10.1093/mnras/sty2794}, \href
  {https://ui.adsabs.harvard.edu/abs/2019MNRAS.482.2166T} {482, 2166}

\bibitem[\protect\citeauthoryear{{Topal}, {Bureau}, {Tiley}, {Davis}  \&
  {Torii}}{{Topal} et~al.}{2018}]{topal2018}
{Topal} S.,  {Bureau} M.,  {Tiley} A.~L.,  {Davis} T.~A.,   {Torii} K.,  2018,
  \mn@doi [\mnras] {10.1093/mnras/sty1617}, \href
  {https://ui.adsabs.harvard.edu/abs/2018MNRAS.479.3319T} {479, 3319}

\bibitem[\protect\citeauthoryear{{Trujillo-Gomez}, {Klypin}, {Primack}  \&
  {Romanowsky}}{{Trujillo-Gomez} et~al.}{2011}]{TG11}
{Trujillo-Gomez} S.,  {Klypin} A.,  {Primack} J.,   {Romanowsky} A.~J.,  2011,
  \mn@doi [\apj] {10.1088/0004-637X/742/1/16}, \href
  {http://adsabs.harvard.edu/abs/2011ApJ...742...16T} {742, 16}

\bibitem[\protect\citeauthoryear{{Tully} \& {Courtois}}{{Tully} \&
  {Courtois}}{2012}]{TC12}
{Tully} R.~B.,  {Courtois} H.~M.,  2012, \mn@doi [\apj]
  {10.1088/0004-637X/749/1/78}, \href
  {http://adsabs.harvard.edu/abs/2012ApJ...749...78T} {749, 78}

\bibitem[\protect\citeauthoryear{{Tully} \& {Fisher}}{{Tully} \&
  {Fisher}}{1977}]{tf77}
{Tully} R.~B.,  {Fisher} J.~R.,  1977, \aap, \href
  {https://ui.adsabs.harvard.edu/abs/1977A&A....54..661T} {500, 105}

\bibitem[\protect\citeauthoryear{{Tully}, {Rizzi}, {Shaya}, {Courtois},
  {Makarov}  \& {Jacobs}}{{Tully} et~al.}{2009}]{tully2009}
{Tully} R.~B.,  {Rizzi} L.,  {Shaya} E.~J.,  {Courtois} H.~M.,  {Makarov}
  D.~I.,   {Jacobs} B.~A.,  2009, \mn@doi [\aj] {10.1088/0004-6256/138/2/323},
  \href {https://ui.adsabs.harvard.edu/abs/2009AJ....138..323T} {138, 323}

\bibitem[\protect\citeauthoryear{{Tully} et~al.,}{{Tully}
  et~al.}{2013}]{tully2013}
{Tully} R.~B.,  et~al., 2013, \mn@doi [\aj] {10.1088/0004-6256/146/4/86}, \href
  {https://ui.adsabs.harvard.edu/abs/2013AJ....146...86T} {146, 86}

\bibitem[\protect\citeauthoryear{{Tully}, {Courtois}, {Hoffman}  \&
  {Pomar{\`e}de}}{{Tully} et~al.}{2014}]{tully2014}
{Tully} R.~B.,  {Courtois} H.,  {Hoffman} Y.,   {Pomar{\`e}de} D.,  2014,
  \mn@doi [\nat] {10.1038/nature13674}, \href
  {https://ui.adsabs.harvard.edu/abs/2014Natur.513...71T} {513, 71}

\bibitem[\protect\citeauthoryear{{Tully}, {Pomar{\`e}de}, {Graziani},
  {Courtois}, {Hoffman}  \& {Shaya}}{{Tully} et~al.}{2019}]{cosmic3}
{Tully} R.~B.,  {Pomar{\`e}de} D.,  {Graziani} R.,  {Courtois} H.~M.,
  {Hoffman} Y.,   {Shaya} E.~J.,  2019, \mn@doi [\apj]
  {10.3847/1538-4357/ab2597}, \href
  {https://ui.adsabs.harvard.edu/abs/2019ApJ...880...24T} {880, 24}

\bibitem[\protect\citeauthoryear{{{\"U}bler} et~al.,}{{{\"U}bler}
  et~al.}{2017}]{ubler2017}
{{\"U}bler} H.,  et~al., 2017, \mn@doi [\apj] {10.3847/1538-4357/aa7558}, \href
  {https://ui.adsabs.harvard.edu/abs/2017ApJ...842..121U} {842, 121}

\bibitem[\protect\citeauthoryear{{Verheijen}}{{Verheijen}}{2001}]{mverheijen2001}
{Verheijen} M. A.~W.,  2001, \mn@doi [\apj] {10.1086/323887}, \href
  {https://ui.adsabs.harvard.edu/abs/2001ApJ...563..694V} {563, 694}

\bibitem[\protect\citeauthoryear{{Verheijen} \& {Sancisi}}{{Verheijen} \&
  {Sancisi}}{2001}]{Verheijen2001}
{Verheijen} M.~A.~W.,  {Sancisi} R.,  2001, \mn@doi [\aap]
  {10.1051/0004-6361:20010090}, \href
  {https://ui.adsabs.harvard.edu/abs/2001A&A...370..765V} {370, 765}

\bibitem[\protect\citeauthoryear{Wang, Koribalski, Serra, van~der Hulst,
  Roychowdhury, Kamphuis  \& N.~Chengalur}{Wang et~al.}{2016}]{wang2016}
Wang J.,  Koribalski B.~S.,  Serra P.,  van~der Hulst T.,  Roychowdhury S.,
  Kamphuis P.,   N.~Chengalur J.,  2016, \mn@doi [MNRAS]
  {10.1093/mnras/stw1099}, 460, 2143–2151

\bibitem[\protect\citeauthoryear{Westmeier, Jurek, Obreschkow, Koribalski  \&
  Staveley-Smith}{Westmeier et~al.}{2013}]{Westmeier2013}
Westmeier T.,  Jurek R.,  Obreschkow D.,  Koribalski B.~S.,   Staveley-Smith
  L.,  2013, \mn@doi [MNRAS] {10.1093/mnras/stt2266}, 438, 1176–1190

\bibitem[\protect\citeauthoryear{{Willick}}{{Willick}}{1994}]{willick94}
{Willick} J.~A.,  1994, \mn@doi [\apjs] {10.1086/191957}, \href
  {http://adsabs.harvard.edu/abs/1994ApJS...92....1W} {92, 1}

\bibitem[\protect\citeauthoryear{{Willick}}{{Willick}}{1999}]{willick99}
{Willick} J.~A.,  1999, \mn@doi [\apj] {10.1086/307108}, \href
  {https://ui.adsabs.harvard.edu/abs/1999ApJ...516...47W} {516, 47}

\bibitem[\protect\citeauthoryear{{den Heijer} et~al.,}{{den Heijer}
  et~al.}{2015}]{Heijer2015}
{den Heijer} M.,  et~al., 2015, \mn@doi [\aap] {10.1051/0004-6361/201526879},
  \href {https://ui.adsabs.harvard.edu/abs/2015A&A...581A..98D} {581, A98}

\makeatother
\end{thebibliography}
\vspace*{1cm}

\end{document}